\documentclass{elsart}
\usepackage{amssymb,amsfonts}
\usepackage{graphicx,color}
\usepackage{bm}

\newcommand{\beq}{\begin{equation}}
\newcommand{\eeq}{\end{equation}}
\newcommand{\beqa}{\begin{eqnarray}}
\newcommand{\eeqa}{\end{eqnarray}}

\newcommand{\dslash}{\!\!\not\!}

\begin{document}
\begin{frontmatter}

\title{Multiple Parton Scattering in Nuclei: Quark-quark Scattering}
\author[label1]{Andreas Sch\"afer,}
\author[label2]{Xin-Nian Wang}
\author[label3,label4,label1]{and Ben-Wei Zhang}

\address[label1]{Institut f\"ur Theoretische Physik, Universit\"at
Regensburg\\ D-93040 Regensburg, Germany}
\address[label2]{Nuclear Science Division, MS 70R0319 \\
Lawrence Berkeley National Laboratory, Berkeley, CA 94720}
\address[label3]{Cyclotron Institute and Physics Department, Texas A$\&$M
University \\ College Station, Texas 77843-3366}
\address[label4]{Institute of Particle Physics,
Central China Normal University \\ Wuhan 430079, China}

\begin{abstract}

Modifications to quark and antiquark fragmentation functions due to
quark-quark (antiquark) double scattering in nuclear medium are
studied systematically up to order $\cal{O}$$(\alpha_s^2)$ in
deeply inelastic scattering (DIS) off nuclear targets. At the order
$\cal{O}$$(\alpha_s^2)$, twist-four contributions from quark-quark
(antiquark) rescattering also exhibit the Landau-Pomeranchuck-Midgal
(LPM) interference feature similar to gluon bremsstrahlung induced
by multiple parton scattering. Compared to quark-gluon scattering,
the modification, which is dominated by $t$-channel quark-quark
(antiquark) scattering, is only smaller by a factor of $C_F/C_A=4/9$
times the ratio of quark and gluon distributions in the medium.
Such a modification is not negligible for realistic kinematics and finite
medium size. The
modifications to quark (antiquark) fragmentation functions from
quark-antiquark annihilation processes are shown to be determined by
the antiquark (quark) distribution density in the medium. The
asymmetry in quark and antiquark distributions in nuclei will
lead to different modifications of quark and antiquark
fragmentation functions inside a nucleus, which qualitatively explains
the experimentally observed flavor dependence of the leading hadron
suppression in semi-inclusive DIS off nuclear targets. The
quark-antiquark annihilation processes also mix quark and gluon
fragmentation functions in the large fractional momentum region,
leading to a flavor dependence of jet quenching in heavy-ion
collisions.

\end{abstract}


\end{frontmatter}

\section{Introduction}

Multiple parton scattering in a dense medium can be used as a useful
tool to study properties of both hot and cold nuclear
matter. The success of such an approach has been demonstrated by the
discovery of strong jet quenching phenomena in central $Au+Au$ collisions
at the Relativistic Heavy-ion Collider
(RHIC) \cite{Adcox:2001jp,Adler:2002xw,starjet} and their
implications on the formation of a strongly coupled quark-gluon
plasma at RHIC \cite{Gyulassy:2004zy,Jacobs:2004qv}. However,
for a convincing phenomenological study of the existing and future
experimental data, a unified description of all medium effects
in hard processes involving nuclei, such as electron-nucleus ($e+A$),
hadron-nucleus ($h+A$) and nucleus-nucleus collisions ($A+A$)
has to be developed \cite{Qiu:2005,Qiu:2001}. This must include the
physics of transverse momentum broadening \cite{Guo:1998}, strong
nuclear enhancement in DIS \cite{Guo:2001tz} and Drell-Yan
production \cite{Fries:2000da,Qiu:2001zj}, nuclear
shadowing \cite{Qiu:2004}, and parton energy loss due to  gluon
radiation induced by multiple scattering \cite{GW1,BDMPS,Zak,GLV,SW,GW,ZW}.

There exist many different frameworks in the literature to describe
multiple scattering in a nuclear medium \cite{kopeli,gvwz,Kovner:2003zj}.
Among them the twist expansion approach is based on the generalized
factorization in perturbative QCD as initially developed by
Luo, Qiu and Sterman (LQS) \cite{LQS}. In the LQS formalism,
multiple scattering processes generally involve high-twist
multiple-parton correlations in analogy to the parton distribution
operators in leading twist processes. Though the corresponding
higher twist corrections are suppressed by powers of $1/Q^2$, they
are enhanced at least by a factor of $A^{1/3}$ due to multiple
scattering in a large nucleus. This framework has been applied recently
to study medium modification of the fragmentation functions as the
leading parton propagates through the medium \cite{GW,ZW}.
Because of the non-Abelian
Landau-Pomeranchuck-Midgal interference in the gluon bremsstrahlung
induced by multiple parton scattering in nuclei, the higher-twist
nuclear modifications to the fragmentation functions are in fact
enhanced by $A^{2/3}$, quadratic in the nuclear size \cite{GW,ZW}.
Phenomenological study of parton energy loss and nuclear modification
of the fragmentation functions in cold nuclear matter  \cite{EW1}
gives a good description of the nuclear modification of the leading
hadron spectra in semi-inclusive deeply inelastic lepton-nucleus
scattering observed by the HERMES experiment \cite{hermes:2000,hermes:2003}.
The same framework also gives a compelling explanation for the suppression
of large transverse momentum hadrons discovered at RHIC \cite{wang03}.

The emphasis of recent studies of medium modification of
fragmentation functions has been on radiative parton energy loss
induced by multiple scattering with gluons. Such processes indeed
are dominant relative to multiple scattering with quarks because of
the abundance of soft gluons in either cold nuclei or hot dense
matter produced in heavy-ion collisions. Since gluon bremsstrahlung
induced by scattering with medium gluons is the same for quarks and
anti-quarks, one also expects the energy loss and fragmentation
modification to be identical for quarks and anti-quarks. However, in
a medium with finite baryon density such as cold nuclei and the forward
region of heavy-ion collisions, the difference between quark and
anti-quark distributions in the medium should lead to different
energy loss and modified fragmentation functions for quarks and
antiquarks through quark-antiquark annihilation processes. To
study such an asymmetry, one must consider systematically all
possible quark-quark and quark-antiquark scattering processes, which
will be the focus of this paper.

In this study we will calculate the modifications of quark and
antiquark fragmentation functions (FF) due to quark-quark
(antiquark) double scattering in a nuclear medium, working within
the LQS framework for generalized factorization in perturbative QCD.
For a complete description of nuclear modification of the single
inclusive hadron spectra, one still have to consider medium
modification of gluon fragmentation functions in addition to
modified quark fragmentation function due to quark-gluon  scattering
[18]. The theoretical results presented in this paper will be a
second step toward a complete description of medium modified
fragmentation functions. However, one can already find that
quark-quark (antiquark) double scattering will give different
corrections to quark and antiquark FF, depending on antiquark and
quark density of the medium, respectively. This difference between
modified quark and antiquark FF may shed light on the interesting
observation by the HERMES experiment \cite{hermes:2000,hermes:2003}
of a large difference between nuclear suppression of the leading
proton and antiproton spectra in semi-inclusive DIS off large
nuclei. Such a picture of quark-quark (antiquark) scattering can
provide a competing mechanism for the experimentally observed
phenomenon in addition to possible absorption of final state hadrons
inside nuclear matter \cite{Falter:2004uc,Kopeliovich:2004}.

The paper is organized as follows. In the next section we will
present the general formalism of our calculation including the
generalized factorization of twist-4 processes. In Section III we
will illustrate the procedure of calculating the hard partonic parts of
quark-quark double scattering in nuclei. In Section IV we will
discuss the modifications to quark and antiquark fragmentation
functions  due to quark-quark (antiquark) double scattering in nuclei.
In Section V, we will focus on the flavor dependent part of the
medium modification to the quark FF's due to quark-antiquark
annihilation and we will discuss the implications for the flavor
dependence of the leading hadron spectra in both DIS off a nucleus
and heavy-ion collisions. We will summarize our work in Section VI.
In the Appendix~\ref{appa}, we collect the complete results for the
hard partonic parts for different cut diagrams of quark-quark (antiquark)
double rescattering in nuclei. We also provide an alternative
calculation of the hard parts of the central-cut diagrams
in Appendix~\ref{appc} through elastic quark-quark scattering or
quark-antiquark annihilation as a cross check.

\section{General formalism}

In order to study quark and antiquark FF's in semi-inclusive
deeply inelastic lepton-nucleus scattering, we consider the
following processes,
$$ e(L_1) + A(p) \longrightarrow e(L_2) + h (\ell_h) +X \ , $$
where $L_1$ and $L_2$ are the four momenta of the incoming and
outgoing leptons, and $\ell_h$ is the observed hadron momentum. The
differential cross section for the semi-inclusive process can be
expressed as
\begin{equation}
E_{L_2}E_{\ell_h}\frac{d\sigma_{\rm DIS}^h}{d^3L_2d^3\ell_h}
=\frac{\alpha^2_{\rm EM}}{2\pi s}\frac{1}{Q^4} L_{\mu\nu}
E_{\ell_h}\frac{dW^{\mu\nu}}{d^3\ell_h} \; ,
\label{sigma}
\end{equation}
where $p = [p^+,0,{\bf 0}_\perp] \label{eq:frame}$
is the momentum per nucleon in the nucleus,
$q =L_2-L_1 = [-Q^2/2q^-, q^-, {\bf 0}_\perp]$ the momentum transfer
carried by the virtual photon,
$s=(p+L_1)^2$ the lepton-nucleon center-of-mass energy
and $\alpha_{\rm EM}$ is the electromagnetic (EM)
coupling constant. The leptonic tensor is given by
$L_{\mu\nu}=1/2\, {\rm Tr}(\gamma \cdot L_1 \gamma_{\mu}
\gamma \cdot L_2 \gamma_{\nu})$
while the semi-inclusive hadronic tensor is defined as,
\begin{eqnarray}
E_{\ell_h}\frac{dW_{\mu\nu}}{d^3\ell_h}&=&
\frac{1}{2}\sum_X \langle A|J_\mu(0)|X,h\rangle
\langle X,h| J_\nu(0)|A\rangle 2\pi \delta^4(q+p-p_X-\ell_h)
\end{eqnarray}
where $\sum_X$ runs over all possible final states and
$J_\mu=\sum_q e_q \bar{\psi}_q \gamma_\mu\psi_q$ is the
hadronic EM current.


\begin{figure}
\begin{center}
\includegraphics[width=80mm]{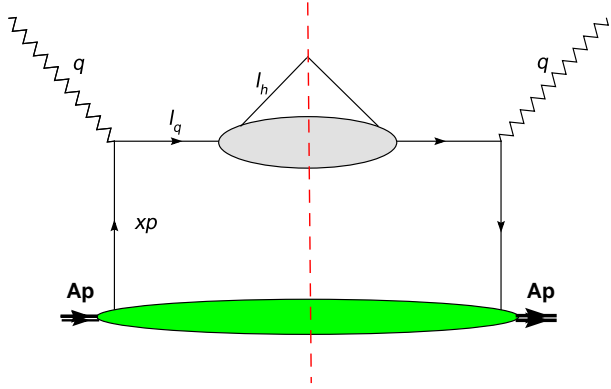}
\end{center}
\caption{Lowest order and leading-twist contribution to
semi-inclusive DIS.}
\label{fig:2q}
\end{figure}

Assuming collinear factorization in the parton model, the
leading-twist contribution to the semi-inclusive cross section can
be factorized into a product of parton distributions, parton
fragmentation functions and the partonic cross section. Including
all leading log radiative corrections, the lowest order contribution
[${\cal O}(\alpha_s^0)$] from a single hard $\gamma^*+ q$ scattering,
as illustrated in Fig.~\ref{fig:2q}, can be written as
\begin{eqnarray}
& &\frac{dW^S_{\mu\nu}}{dz_h}
= \sum_q \int dx f_q^A(x,\mu_I^2) H^{(0)}_{\mu\nu}(x,p,q)
D_{q\to h}(z_h,\mu^2)\, ; \label{Dq} \\
& &H^{(0)}_{\mu\nu}(x,p,q) = \frac{e_q^2}{2}\, {\rm Tr}(\gamma \cdot p
\gamma_{\mu} \gamma \cdot(q+xp) \gamma_{\nu}) \, \frac{2\pi}{2p\cdot
q} \delta(x-x_B) \, , \label{H0}
\end{eqnarray}
where the momentum fraction carried by the hadron is defined as
$z_h=\ell_h^-/q^-$, $x_B=Q^2/2p^+q^-$ is the Bjorken scaling variable,
$\mu_I^2$ and $\mu^2$ are the factorization scales for the initial
quark distributions $f_q^A(x,\mu_I^2)$ in a nucleus and the
fragmentation functions in vacuum $D_{q\to h}(z_h,\mu^2)$, respectively. The
renormalized quark fragmentation function $D_{q\to h}(z_h,\mu^2)$
satisfies the DGLAP QCD evolution equations \cite{AP}:
\begin{eqnarray}
  \frac{\partial D_{q\to h}(z_h,\mu^2)}{\partial \ln \mu^2} & = &
  \frac{\alpha_s(\mu^2)}{2\pi} \int^1_{z_h} \frac{dz}{z}
\left[ \gamma_{q\to qg}(z) D_{q\to h}(z_h/z,\mu^2) \right. \nonumber \\
& & \hspace{1.5in}+ \left. \gamma_{q\to gq}(z)
D_{g\to h}(z_h/z,\mu^2)\right];  \label{eq:ap1} \\
\frac{\partial D_{g\to h}(z_h,\mu^2)}{\partial \ln \mu^2} &
= & \frac{\alpha_s(\mu^2)}{2\pi} \int^1_{z_h} \frac{dz}{z} \left[
    \sum_{q=1}^{2n_f} \gamma_{g\to q\bar{q}}(z)
  D_{q\to h}(z_h/z,\mu^2) \right. \nonumber \\
& & \hspace{1.5in} + \left. \gamma_{g\to gg}(z)
 D_{g\to h}(z_h/z,\mu^2)\right] , \label{eq:ap2}
\end{eqnarray}
where $\gamma_{a\to bc}(z)$ denotes the splitting functions
of the corresponding radiative processes \cite{field,Peskin}.

In DIS off a nuclear target, the propagating quark will experience
additional scatterings with other partons from the nucleus. The
rescatterings may induce additional parton (quark or gluon) radiation
and cause the leading quark to lose energy. Such induced  radiation
will effectively give rise to additional terms in the DGLAP
evolution equation leading to a modification of the fragmentation
functions in a medium. These are power-suppressed higher-twist
corrections and they involve higher-twist parton matrix elements.
We will only consider those contributions that involve two-parton
correlations from two different nucleons inside the nucleus. They
are proportional to the thickness of the nucleus \cite{GW,LQS,OW} and
thus are enhanced by a nuclear factor $A^{1/3}$ as compared to
two-parton correlations in a nucleon. As in previous studies
\cite{GW,ZW}, we will limit our study to such double scattering
processes in a nuclear medium. These are twist-four processes
and give leading contributions to the nuclear effects.
The contributions of higher twist processes or contributions not
enhanced by the nuclear medium will be neglected for the time being.

\begin{figure}
\begin{center}
\includegraphics[width=80mm]{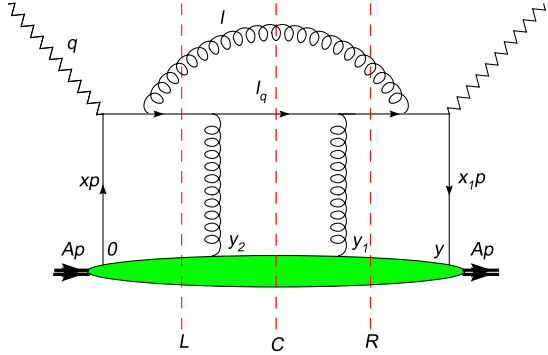}
\end{center}
\caption{A typical diagram for quark-gluon double scattering  with
three possible cuts [central(C), left(L) and right(R)].}
\label{fig:qg}
\end{figure}

When considering double scattering with nuclear enhancement, a very
important process is quark-gluon double scattering as
illustrated in Fig.~\ref{fig:qg}. Such processes give the dominant
contribution to the leading quark energy loss and have been studied
in detailed in Refs. \cite{GW,ZW}. The modification to the vacuum
quark fragmentation function from quark-gluon scattering is,
\begin{eqnarray}
\Delta D^{qg\rightarrow qg}_{q\rightarrow h}(z_h)
&=&\frac{\alpha_s^2 C_A}{N_c} \int  \frac{d\ell_T^2}{\ell_T^4}
\int_{z_h}^1\frac{dz}{z} \left\{
D_{q\rightarrow h}(z_h/z)  \left[
\frac{1+z^2}{(1-z)_+}\frac{T^A_{qg}(x,x_L)}{f_q^A(x)}
\right. \right. \nonumber \\
& &  \hspace{-0.9in}+ \left. \left. \delta(z-1)
\frac{\Delta T^A_{qg}(x,\ell_T^2)}{f_q^A(x)} \right]
+D_{g\rightarrow h}(z_h/z)\left[
\frac{1+(1-z)^2}{z}\frac{T^A_{qg}(x,x_L)}{f_q^A(x)}\right] \right \},
\label{eq:dd-qg}
\end{eqnarray}
where the $+$function is defined as
\begin{equation}
\int_0^1 dz \frac{F(z)}{(1-z)_+} \equiv \int_0^1 dz \frac{F(z)-F(1)}{1-z}
\label{eq:plus}
\end{equation}
for any $F(z)$ that is sufficiently smooth at $z=1$ and the twist-four
quark-gluon correlation function,
\begin{eqnarray}
T^A_{qg}(x,x_L)&=&\int \frac{dy^-}{2\pi}dy_1^-dy_2^-
e^{i(x+x_L)p^+y^-}(1-e^{-ix_Lp^+y_2^-})
(1-e^{-ix_Lp^+(y^--y_1^-)}) \nonumber \\
&&\hspace{-0.4in}\times \langle A | \bar{\psi}_q(0)\,
\frac{\gamma^+}{2}F_{\sigma}^{\ +}(y_{2}^{-}) F^{+\sigma}(y_1^{-})
\psi_q(y^{-})| A\rangle \theta(-y_2^-)\theta(y^- -y_1^-),
\label{eq:qgmatrix}
\end{eqnarray}
has explicit interference included. The matrix element in the virtual
correction [the term with $\delta (z-1)$] is defined as
\begin{eqnarray}
\Delta T^A_{qg}(x,\ell_T^2) \equiv
\int_0^1 dz\frac{1}{1-z}\left[ 2 T^A_{qg}(x,x_L)|_{z=1}
-(1+z^2) T^A_{qg}(x,x_L)\right] \, . \label{eq:vsplit}
\end{eqnarray}
Since $T^A_{qg}(x,x_L)/f_q^A(x)$ is proportional to gluon distribution
and independent of the flavor of the leading quark, the suppression of
the hadron spectrum caused by quark-gluon or antiquark-gluon
scattering should be proportional to the gluon density of the medium and
is identical for quark and antiquark fragmentation.
It was shown in Ref.~\cite{EW1}
that such modification of parton fragmentation functions by quark-gluon
double scattering and gluon bremsstrahlung in a nuclear medium  describes
very well the recent HERMES data \cite{hermes:2000} on semi-inclusive
DIS off nuclear targets.

\begin{figure}
\begin{center}
\includegraphics[width=80mm]{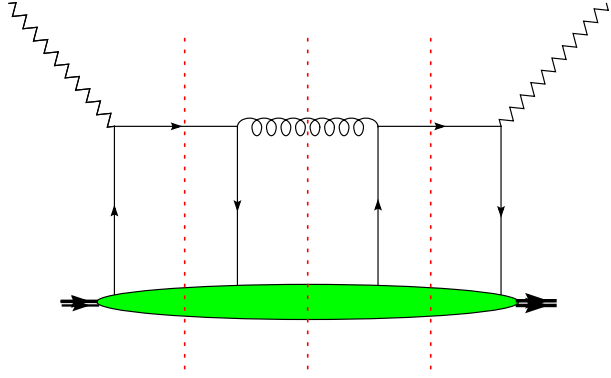}
\end{center}
\caption{Diagram for leading order
quark-antiquark annihilation with three possible cuts [central(C),
left(L) and right(R)].}
\label{fig0}
\end{figure}

\begin{figure}
\begin{center}
\includegraphics[width=80mm]{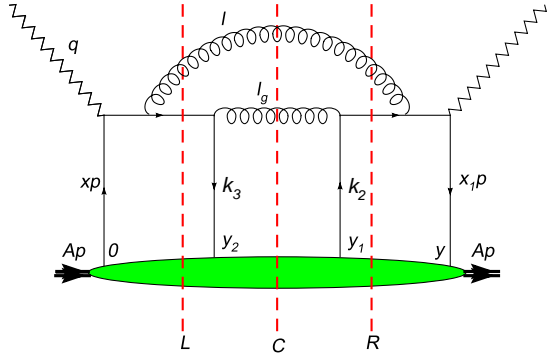}
\end{center}
\caption{A typical diagram for next-to-leading order correction to
quark-antiquark annihilation with three possible cuts [central(C),
left(L) and right(R)].}
\label{fig:4q-EX}
\end{figure}

In this paper, we will consider quark-quark (antiquark) double
scattering such as the process shown in Fig.~\ref{fig0} and
its radiative corrections at order
$\cal{O}$$(\alpha_s^2)$ in Fig.~\ref{fig:4q-EX}.
The contributions of quark-quark double scattering is proportional to
the quark density in a nucleon, while the contribution
of quark-gluon double scattering is proportional to the gluon density in
a nucleon; and the gluon density is generally larger than the quark
density in a nucleon at small momentum fraction.
However, as pointed out in earlier works \cite{GW}, quark-quark double
scattering mixes quark and gluon fragmentation functions and therefore
gives rise to new nuclear effects. The annihilation processes as
shown in Figs.~\ref{fig0} and \ref{fig:4q-EX} will lead to different
modifications of quark and antiquark fragmentation functions
in a medium with finite baryon
density (or valence quarks). Such differences will in turn lead to
flavor dependence of the nuclear modification of leading
hadron spectra as observed in HERMES experiment
\cite{hermes:2000,hermes:2003}.

Quark-quark double scattering as well as quark-gluon double
scattering are twist-4 processes. We will apply the same
generalized factorization procedure for twist-4 processes as developed
by LQS \cite{LQS} for semi-inclusive processes in DIS. In general, the
twist-four contributions can be expressed as the convolution of
partonic hard parts and two-parton correlation matrix elements. In
this framework, contributions from double quark-quark scattering
in any order of $\alpha_s$, {\it e.g.}, the quark-antiquark annihilation
process as illustrated in Fig.~\ref{fig:4q-EX}, can be written in the
following form,
\begin{eqnarray}
\frac{dW_{\mu\nu}^D}{dz_h} &=& \sum_q
\int \frac{p^+dy^-}{2\pi}\, dy_1^-\, dy_2^-\,
 \overline{H}^D_{\mu\nu}(y^-,y_1^-,y^-_2,p,q,z_h) \nonumber \\
&&\hspace{1.in}\times \langle A|\bar{\psi}_q(0)\frac{\gamma^+}{2}\psi_{q}(y^-)
\bar{\psi}_q(y_1^-)\frac{\gamma^+}{2}\psi_{q}(y_2^-)|A\rangle.
\label{factorize}
\end{eqnarray}
Here we have neglected transverse momenta of all quarks in the hard
partonic part. Transverse momentum dependent contributions are higher twist
and are suppressed by $\langle k_{\perp}^2\rangle /Q^2$, Therefore, all
quarks' momenta are assumed collinear, $k_2=x_2p$ and $k_3=x_3p$.
$\overline{H}^D_{\mu\nu}(y^-,y_1^-,y^-_2,p,q,z_h)$ is the Fourier
transform of the partonic hard part
$\widetilde{H}_{\mu\nu}(x,x_1,x_2,p,q,z_h)$ in momentum space,
\begin{eqnarray}
\overline{H}^D_{\mu\nu}(y^-,y_1^-,y^-_2,p,q,z_h) &=&
\int dx\,\frac{dx_1}{2\pi}\, \frac{dx_2}{2\pi}\,
       e^{ix_1p^+y^- + ix_2p^+y_1^- + i(x-x_1-x_2)p^+y_2^-}
\nonumber \\
&&\hspace{1.0in}\times\widetilde{H}^D_{\mu\nu}(x,x_1,x_2,p,q,z_h)\
\nonumber \\
&=&\int dx H_{\mu\nu}^{(0)}(x,p,q) \overline{H}^D(y^-,y_1^-,y^-_2,x,p,q,z_h)
 \, , \label{eq-mH}
\end{eqnarray}
where, in collinear approximation, the hard partonic
part $H_{\mu\nu}^{(0)}(x,p,q)$ [Eq.~(\ref{H0})] in the leading
twist without multiple parton scattering can be factorized
out of the high-twist hard part $\widetilde{H}^D_{\mu\nu}(x,x_1,x_2,p,q,z_h)$.
The momentum fractions $x, x_1,$ and $x_2$ are fixed by $\delta$-functions
of the on-shell conditions of the final state partons and poles of parton
propagators in the partonic hard part. The phase factors
in $\overline{H}^D_{\mu\nu}(y^-,y_1^-,y^-_2,p,q,z_h)$
can then be factored out, which
in turn will be combined with the partonic fields in Eq.~(\ref{factorize})
to give twist-four partonic matrix elements or two-parton correlations.
The quark-quark double scattering corrections in Eq.~(\ref{factorize})
can then be factorized as the convolution of fragmentation functions,
twist-four partonic matrix elements and the partonic hard scattering
cross sections. For scatterings (versus the annihilation) with
quarks (antiquarks), a summation over the flavor of the secondary
quarks (antiquarks) should be included in two-quark correlation
matrix elements and both $t$, $u$ channels and their interferences
should be considered for scattering of identical quarks in the hard
partonic parts.

After factorization, we then define the twist-four correction to the
leading twist quark fragmentation function in the same form [Eq.~(\ref{Dq})],
\begin{equation}
\frac{dW_{\mu\nu}^D}{dz_h} \equiv
\sum_q \int dx f_q^A(x) H^{(0)}_{\mu\nu}(x,p,q)
\Delta D_{q\rightarrow h}(z_h) \, . \label{eq:qq-dd}
\end{equation}

\section{Quark-quark double scattering processes}
\label{seciii}

In this section we will discuss the calculation of the hard part of
quark-quark double scattering in detail.
The lowest order process of quark-quark (antiquark)
double scattering in nuclei is
quark-antiquark annihilation (or quark-gluon conversion) as shown in
Fig.~\ref{fig0}. The hard partonic parts from the three cut
diagrams in this figure are \cite{GW}:
\begin{eqnarray}
\overline{H}^D_{0,C}(y^-,y_1^-,y_2^-,x,p,q,z_h)&=& D_{g\to h}(z_h)
\frac{2\pi\alpha_s}{N_c}2C_F\frac{x_B}{Q^2} \,
e^{ixp^+y^-} \nonumber \\ &&\hspace{1.0in}\times
 \theta(-y_2^-)\theta(y^- - y_1^-) \, , \label{eq:H-0-C} \\
\overline{H}^D_{0,L}(y^-,y_1^-,y_2^-,x,p,q,z_h)&=& D_{q\to h}(z_h)
\frac{2\pi\alpha_s}{N_c}2C_F\frac{x_B}{Q^2} \, e^{ixp^+y^-}
 \nonumber \\ &&\hspace{1.0in}\times
\theta(y_1^- - y_2^-)\theta(y^- - y_1^-)  \, , \label{eq:H-0-L} \\
\overline{H}^D_{0,R}(y^-,y_1^-,y_2^-,x,p,q,z_h)&=& D_{q\to h}(z_h)
\frac{2\pi\alpha_s}{N_c}2C_F\frac{x_B}{Q^2} \, e^{ixp^+y^-}
\nonumber \\ &&\hspace{1.0in}\times
\theta(-y_2^-)\theta(y_2^- - y_1^-) \, \, . \label{eq:H-0-R}
\end{eqnarray}

The main focus of this paper is about contributions from the
next-leading order corrections to the above lowest order process.
There is a total of 12 diagrams for real corrections at one-loop level as
illustrated in Fig.~\ref{fig1} to Fig.~\ref{fig8}  in the Appendix~\ref{appa},
each having up to three different cuts. In this section, we
demonstrate the calculation of the hard parts from the quark-antiquark
annihilation in Fig.~\ref{fig:4q-EX} in detail as an example. We will
list the complete results of all diagrams in Appendix A.

One can write down the hard partonic part of the central-cut diagram
of Fig.~\ref{fig:4q-EX} (Fig.~\ref{fig1} in Appendix~\ref{appa})
according to the conventional Feynman rule,
\begin{eqnarray}
\overline{H}^D_{C\,\mu\nu}(y^-,y_1^-,y_2^-,p,q,z_h)&=&
\int_{z_h}^1\frac{dz}{z} D_{g\to h}(\frac{z_h}{z})
\int dx\frac{dx_1}{2\pi}\frac{dx_2}{2\pi} e^{ix_1p^+y^- + ix_2p^+y_1^-}
\nonumber \\
&&\hspace{-1.9in}\times e^{i(x-x_1-x_2)p^+y_2^-}
 \int \frac{d^4\ell}{(2\pi)^4} {\rm
Tr}\left[\frac{\!\!\not\!p}{2} \gamma_\mu \widehat{H} \gamma_\nu
\right] 2\pi\delta_+(\ell^2) 2\pi\delta_+(\ell_g^2)\,
\delta(1-z-\frac{\ell^-}{q^-}) \; , \nonumber \\
&&\hspace{-1.5in}\widehat{H} = \frac{C_F^2}{N_c}g^4\frac{\gamma\cdot(q+x_1
p)}{(q+x_1p)^2-i\epsilon} \,\gamma_\alpha\,\frac{\gamma\cdot(q + x_1
p -\ell)}{(q+x_1p-\ell)^2-i\epsilon} \,\gamma_\beta \,
\frac{\dslash{p}}{2}\, \gamma_{\sigma}\nonumber \\
&&\hspace{-1.0in}\times \frac{\gamma\cdot(q + x p
-\ell)}{(q+x p-\ell)^2+i\epsilon}
\gamma_{\rho}\,
 \frac{\gamma\cdot(q+x p)}{(q+x p)^2+i\epsilon} \,
 \varepsilon^{\alpha\rho}(\ell)\,
\varepsilon^{\beta\sigma}(\ell_g) \, \,  ,
 \label{eq:fig-Ex-2}
\end{eqnarray}
where $\delta_+$ is a Dirac $delta$-function with only the
positive solution in its functional variable,
$ \varepsilon^{\alpha\rho}(\ell)=-g^{\alpha\rho} +
(n^\alpha\ell^\rho+n^\rho\ell^\alpha)/n\cdot\ell$
is the polarization tensor of a gluon propagator in an axial gauge
($n\cdot A=0$) with $n=[1,0^-,\vec{0}_\perp]$, $\ell$ and
$\ell_g=q+(x_1+x_2)p-\ell$ are the 4-momenta carried by the two final
gluons respectively. The fragmenting gluon carries a fraction,
$z=\ell_g^-/q^-$, of the initial quark's longitudinal momentum
(the large minus component).

To simplify the calculation in the case of small transverse
momentum $\ell_T\ll q^-, p^+$, we can apply the collinear
approximation to complete the trace of the product of
$\gamma$-matrices,
\begin{equation}
\widehat{H} \approx \gamma\cdot{\ell_q} \frac{1}{4\ell_g^-} {\rm Tr}
\left[\gamma\cdot{\ell_q} \widehat{H} \right] \, .
\label{coll}
\end{equation}
According to the convention in Eqs.~(\ref{factorize}) and (\ref{eq-mH}),
contributions from quark-quark double scattering in the nuclear medium
to the semi-inclusive hadronic tensor in DIS off a nucleus can be expressed
in the general factorized form:
\begin{eqnarray}
\frac{dW^{D}_{q\bar q,\mu\nu}}{dz_h}&=&\sum_q \int dx
H^{(0)}_{\mu\nu}(x,p,q)
\int\frac{p^+ dy^-}{2\pi}dy_1^-dy_2^-
\overline{H}^D(y^-,y_1^-,y_2^-,x,p,q,z_h) \nonumber \\
&\times& \langle A|\bar{\psi}_q(0)\frac{\gamma^+}{2}\psi_{\bar q}(y^-)
\bar{\psi}_q(y_1^-)\frac{\gamma^+}{2}\psi_{\bar q}(y_2^-)|A\rangle\, .
\label{eq:genal}
\end{eqnarray}
After carrying out the momentum integration in $x$, $x_1$, $x_2$ and
$\ell^{\pm}$ in Eq.~(\ref{eq:fig-Ex-2})
with the help of contour integration and $\delta$-functions,
one obtains the hard partonic part, $\overline{H}^D$, of the rescattering
for the central-cut diagram in Fig.~\ref{fig:4q-EX} (Fig.~\ref{fig1}) as
\begin{eqnarray}
\overline{H}^D_{\ref{fig1},C}(y^-,y_1^-,y_2^-,x,p,q,z_h)&=&
 \frac{\alpha_s^2 x_B}{Q^2} \int\frac{d\ell_T^2}{\ell_T^2}\,
\int_{z_h}^1\frac{dz}{z} D_{g\to h}(z_h/z) \frac{C_F^2}{N_c}
\nonumber \\
&\times&  \frac{2(1+z^2)}{z(1-z)}
\overline{I}_{\ref{fig1},C}(y^-,y_1^-,y_2^-,x,x_L,p)
 \, , \label{eq:HEC} \\
\overline{I}_{\ref{fig1},C}(y^-,y_1^-,y_2^-,x,x_L,p)
&=&e^{i(x+x_L)p^+y^-}
\theta(-y_2^-)\theta(y^- - y_1^-) \nonumber \\
&\times &(1-e^{-ix_Lp^+y_2^-})(1-e^{-ix_Lp^+(y^- - y_1^-)}) \, ,
\label{eq:IEC}
\end{eqnarray}
where the momentum fractions $x_L$ is defined as
\begin{equation}
  x_L=\frac{\ell_T^2}{2p^+q^-z(1-z)} \,\,   .
\label{xld}
\end{equation}
Note that the function $\overline{I}_{\ref{fig1},C}(y^-,y_1^-,y_2^-,x,x_L,p)$
contains only phase factors. One can combine these phase factors
with the matrix elements of the quark fields to define a special
two-quark correlation function
\begin{eqnarray}
T_{q\bar q}^{A(\ref{fig1},C)}(x,x_L)&=&\int\frac{p^+ dy^-}{2\pi}dy_1^-dy_2^-
\langle A|\bar{\psi}_q(0)\frac{\gamma^+}{2}\psi_{\bar q}(y^-)
\bar{\psi}_q(y_1^-)\frac{\gamma^+}{2}\psi_{\bar q}(y_2^-)|A\rangle \nonumber \\
&\times& \overline{I}_{\ref{fig1},C}(y^-,y_1^-,y_2^-,x,x_L,p)\, .
\end{eqnarray}
The contribution from quark-antiquark annihilation in
the central-cut diagram in Fig.~\ref{fig:4q-EX} to the hadronic tensor
can then be expressed as
\begin{eqnarray}
\frac{dW^{D}_{q\bar q,\mu\nu}}{dz_h}&=&\sum_q \int dx
H^{(0)}_{\mu\nu}(x,p,q) \frac{\alpha_s^2 x_B}{Q^2}
\int\frac{d\ell_T^2}{\ell_T^2}\, \int_{z_h}^1\frac{dz}{z}
D_{g\to h}(\frac{z_h}{z}) \nonumber \\
&&\hspace{1.5in} \times\frac{C_F^2}{N_c}  \frac{2(1+z^2)}{z(1-z)}
 T_{q\bar q}^{A(\ref{fig1},C)}(x,x_L).
\label{eq:qqbar-fac}
\end{eqnarray}
Contributions from all quark-quark (antiquark) double scattering
processes can be cast in the above factorized form.

The structure of the phase factors
in $\overline{I}_{\ref{fig1},C}(y^-,y_1^-,y_2^-,x,x_L,p)$ is exactly the same
as for gluon bremsstrahlung induced by quark-gluon scattering as
studied in Ref.~\cite{GW,ZW}. It resembles the cross section of dipole
scattering and represents contributions from two different processes
and their interferences. It contains essentially four terms,
\begin{eqnarray}
\overline{I}_{\ref{fig1},C}(y^-,y_1^-,y_2^-,x,x_L,p)&=&
   \theta(-y_2^-)\theta(y^- - y_1^-)e^{i(x+x_L)p^+y^-}\nonumber \\
   & & \hspace{-1.0in}\times[1+e^{-ix_Lp^+(y^- + y_2^- - y_1^-)}
   -e^{-ix_Lp^+y_2^-}-e^{-ix_Lp^+(y^- - y_1^-)}] \, .
\end{eqnarray}
The first term corresponds to the so-called hard-soft processes
where the gluon emission is induced by the hard scattering
between the virtual photon $\gamma^*$ and the initial quark
with momentum $(x+x_L)p$. The quark then becomes on-shell before
it annihilates with a soft antiquark from the nucleus that carries
zero momentum and converts into a real gluon in the final state.
The second term corresponds to a process in which the initial
quark with momentum $xp$ is on-shell after the first
hard $\gamma^*$-quark scattering. It then annihilates with another
antiquark and produces two final gluons in the final state. In this
process, the antiquark carries finite (hard) momentum $x_Lp$.
Therefore one often refers to this process as double-hard scattering
as compared to the first process in which the antiquark carries zero
momentum. Set aside the change of flavors in the initial and final
states, the double-hard scattering corresponds essentially to
two-parton elastic scattering with finite momentum and energy
transfer. This is in contrast to the hard-soft scattering
which is essentially the final state radiation of the $\gamma^*$-quark
scattering and the total energy and momentum of the two final state
gluons all come from the initial quark. The corresponding
matrix elements of the two-quark correlation functions from these
first two terms are called `diagonal' elements.

The third and fourth terms with negative signs in
$\overline{I}_{\ref{fig1},C}(y^-,y_1^-,y_2^-,x,x_L,p)$
are interferences between hard-soft and double hard processes.
The corresponding matrix elements are called `off-diagonal'.
The cancellation between the two diagonal and off-diagonal terms
essentially gives rise to the destructive interference which
is very similar to the Landau-Pomeranchuk-Migdal (LPM)
interference in gluon bremsstrahlung induced by quark-gluon
double scattering \cite{GW,ZW}. One can similarly define the
formation time of the parton (quark or gluon) emission as
\begin{equation}
\tau_f \equiv \frac{1}{x_Lp^+} \,\,\, .
\end{equation}
In the limit of collinear emission ($x_L\to 0$) or when the
formation time of the parton emission, $\tau_f$, is much larger than
the nuclear size, the effective matrix element vanishes because
\begin{equation}
\overline{I}_{\ref{fig1},C}(y^-,y_1^-,y_2^-,x,x_L,p)|_{x_L=0}\to 0 \,\,
, \label{eq:LPM}
\end{equation}
when the hard-soft and double hard processes have complete destructive
interference.

We should note that in the central-cut diagram of Fig.~\ref{fig:4q-EX},
the final state partons are two gluons. Therefore, in Eq.~(\ref{eq:HEC})
the gluon fragmentation function in vacuum $D_{g\to h}(z_h/z)$ enters.
If the other gluon (close to the $\gamma^*$-quark interaction)
fragments, the contribution to the semi-inclusive
hadronic tensor is similar except that the corresponding
effective ``splitting function'' should be replaced by
\begin{equation}
\frac{1+z^2}{z(1-z)} \rightarrow \frac{1+(1-z)^2}{z(1-z)}\, .
\label{eq:split-rep}
\end{equation}
As we will show in Appendix~\ref{appa}, the two gluons in the
quark-antiquark annihilation processes (central-cut diagrams)
are symmetric when
contributions from all possible annihilation processes and their
interferences are summed. Therefore, one can
simply multiply the final results by a factor of 2 to take into
account the hadronization of the second final-state gluon.

In addition to the central-cut diagram, one should also take into
account the asymmetrical-cut diagrams in Fig.~\ref{fig:4q-EX}, which
represent interference between gluon emission from single and triple
scattering.  The hard partonic parts are mainly the same as for the
central-cut diagram. The only differences are in the phase factors
and the fragmentation functions since the fragmenting parton can be
the final-state quark or gluon. These hard parts can be calculated
following  a similar procedure and one gets,
\begin{eqnarray}
\overline{H}^D_{\ref{fig1},L(R)}(y^-,y_1^-,y_2^-,x,p,q,z_h)&=&
\frac{\alpha_s^2 x_B}{Q^2} \int\frac{d\ell_T^2}{\ell_T^2}
\int_{z_h}^1\frac{dz}{z} D_{q\to h}(\frac{z_h}{z})
\frac{C_F^2}{N_c} \frac{2(1+z^2)}{z(1-z)} \nonumber \\
&\times&
\overline{I}_{\ref{fig1},L(R)}(y^-,y_1^-,y_2^-,x,x_L, p)
 \, , \label{eq:EX-LR} \\
\overline{I}_{\ref{fig1},L}(y^-,y_1^-,y_2^-,x,x_L,p)\,
&=&-e^{i(x+x_L)p^+y^- }(1-e^{-ix_Lp^+(y^- - y_1^-)})
\nonumber \\
&&\hspace{0.5in}\times
\theta(y_1^- - y_2^-)\theta(y^- - y_1^-) \, ,\label{eq:I-EX-L} \\
\overline{I}_{\ref{fig1},R}(y^-,y_1^-,y_2^-,x,x_L,p)
&=&-e^{i(x+x_L)p^+y^-}(1-e^{-ix_Lp^+y_2^-})
\nonumber \\ &&\hspace{0.5in} \times
\theta(-y_2^-)\theta(y_2^- - y_1^-)\, .\label{eq:I-EX-R}
\end{eqnarray}
In the asymmetrical cut diagrams, the above contributions come
from the fragmentation of the final-state quark. Therefore,
quark fragmentation function $D_{q\to h}(z_h/z)$ enters this
contribution. For gluon fragmentation into the observed hadron in
this asymmetrical-cut diagrams, the contribution
can be obtained by simply replacing the quark fragmentation function
by the gluon fragmentation function $D_{g\to h}(z_h/z)$
and replacing $z$ by $1-z$.  Summing the contributions from
three different cut diagrams of
Fig.~\ref{fig:4q-EX}, we can observe further examples of
mixing (or conversion) of quark and gluon fragmentation
functions. This medium-induced mixing was first observed by Wang and
Guo \cite{GW} and is a unique feature of quark-quark (antiquark)
double scattering among all multiple parton scattering processes.

With the same procedure we can calculate contributions from all other
cut diagrams of quark-quark (antiquark)
double scattering at order $\cal{O}$$(\alpha_s^2)$,
which are listed in Appendix~\ref{appa}. There are three types of
processes: two annihilation processes, $q\bar q\rightarrow gg$
(central-cut diagrams in Figs.~\ref{fig1}, \ref{fig5}, \ref{fig14}, \ref{fig9}
and \ref{fig10}), $q\bar q\rightarrow q_i\bar q_i$ (central-cut
diagram in Fig.~\ref{fig13}) and quark-quark (antiquark)
scattering, $q q_i(\bar q_i)\rightarrow q q_i(\bar q_i)$ (central-cut
diagram in  Fig.~\ref{fig2}). One
also has to consider the interference of $s$ and $t$-channel
amplitude for annihilation into an identical quark pair,
$q\bar q\rightarrow q\bar q$ (central-cut diagrams in Figs.~\ref{fig3}
and \ref{fig4}) and the
interference between $t$ and $u$ channels of identical quark
scattering $q q\rightarrow q q$ (central-cut diagram in Fig.~\ref{fig21}).

Contributions from left and right-cut diagrams correspond to interference
between the amplitude of gluon radiation from single $\gamma*$-quark
scattering and triple quark scattering. The amplitudes of
gluon radiation via triple quark scattering essentially come from
radiative corrections to the left and right-cut diagrams of the
lowest-order quark-antiquark annihilation in Fig.~\ref{fig0}
(as shown in left and right-cut diagrams in Figs.~\ref{fig1}, \ref{fig5},
\ref{fig9}, \ref{fig10}, \ref{fig3}, \ref{fig4}, \ref{fig7}
and \ref{fig8}). Two other
triple quark scatterings with gluon radiation, shown as the left
and right-cut diagrams in Figs.~\ref{fig2} and \ref{fig21}, correspond
to the case where one of the final state quarks, after quark-quark scattering,
annihilates with another antiquark and converts into a final state gluon.

\section{Modified Fragmentation Functions}
\label{MFF}

In order to simplify the contributions from quark-quark (antiquark)
scattering (annihilation), one can first organize the results of the hard
parts in terms of contributions from central, left or right-cut diagrams,
which are associated by contour integrals with specific products of
$\theta$-functions,
\begin{eqnarray}
\overline{H}^D=H^D_{C} \theta(-y_2^-)\theta(y^- - y_1^-)
&-&H^D_{L}\theta(y_1^- - y_2^-)\theta(y^- - y_1^-) \nonumber \\
& & \hspace{0.5in}- H^D_{R}\theta(-y_2^-)\theta(y_2^- - y_1^-) \, .
\label{eq:contact}
\end{eqnarray}
These $\theta$-functions provide a space-time ordering of
the parton correlation and will restrict the
integration range along the light-cone. For contributions from central,
left and right-cut diagrams that have identical hard partonic parts,
$H^{D(c)}_{C}=H^{D(c)}_{L}=H^{D(c)}_{R}$,
they will have a common combination of $\theta$-functions
that produces a path-ordered integral,
\begin{eqnarray}
\int_0^{y^-}dy_1^-\int_0^{y_1^-}dy_2^- &=&
-\int dy_1^-dy_2^-\left[\theta(-y_2^-)\theta(y^--y_1^-)
-\theta(-y_2^-)\theta(y_2^--y_1^-) \right. \nonumber \\
&& \hspace{1.5in} \left. -\theta(y^--y_1^-)\theta(y_1^--y_2^-)
\right]
\label{eq:theta}
\end{eqnarray}
that is limited only by the spatial-spread $y^-$ of the first parton along
the light-cone coordinate. For a high-energy parton that carries
momentum fraction $xp^+$, $y^-\sim 1/xp^+$ should be very small.
Those contributions that are proportional to the above path-ordered
integral are referred to as contact contributions
(or contact interactions).

Similarly, $y_1^- - y_2^-$ is the spatial spread of the second
parton and can only be limited by the spatial size of its host
nucleon even for small value of momentum fraction. The spatial
position of its host nucleon, $y_1^- + y_2^-$, however, can be
anywhere within the nucleus. Therefore, any contributions from
double parton scattering that have unrestricted integration
over $y_1^-$ and $y_2^-$ should be proportional to the nuclear
size of the target $A^{1/3}$ and therefore are nuclear enhanced.
In this paper, we will only keep the nuclear enhanced contributions
and neglect the contact contributions. This will greatly simplify
the final results for double parton scattering.

\subsection{ $q\bar q \rightarrow g$ annihilation}

For the lowest order of quark-antiquark annihilation in
Eqs.~(\ref{eq:H-0-C})-(\ref{eq:H-0-R}), the hard parts from
the three cut diagrams are almost the same except for the parton
fragmentation functions. The central-cut diagram is proportional
to the gluon fragmentation function while the left and right-cut diagrams
are proportional to quark fragmentation functions. Rearranging the
contributions from the three cut diagrams and neglecting the contact term
that is proportional to the path-ordered integral as in
Eq.~(\ref{eq:theta}), the total contribution can be written as
\begin{eqnarray}
\frac{dW^{D(0)}_{\mu\nu}}{dz_h}&=&\sum_q \int dx T^{A(H)}_{q \bar q}(x,0)
\frac{2\pi\alpha_s}{N_c} 2C_F\frac{x_B}{Q^2} H^{(0)}_{\mu\nu}(x,p,q)
\nonumber \\
&&\hspace{2.0in}\times
\left[D_{g\to h}(z_h)-D_{q\to h}(z_h) \right] \, . \label{eq:qq-0}
\end{eqnarray}
According to our definition in
Eq.~(\ref{eq:qq-dd}) of the twist-four correction to the quark
fragmentation functions, the modification
to the quark fragmentation function from the lowest order quark-antiquark
annihilation is then,
\begin{equation}
\Delta D_{q\rightarrow h}^{(q\bar q\rightarrow g)}(z_h)=
\frac{2\pi\alpha_s}{N_c} 2C_F\frac{x_B}{Q^2}
\left[D_{g\to h}(z_h)-D_{q\to h}(z_h) \right]
\frac{T^{A(H)}_{q \bar q}(x,0)}{f^A_q(x)} \, .
\label{eq:dd-qq-g}
\end{equation}
Here the effective quark-antiquark correlation function
$T^{A(H)}_{q{\bar q}}(x,0)$ is defined as,
\begin{eqnarray}
T^{A(H)}_{q{\bar q_i}}(x,x_L)&\equiv&\int\frac{p^+dy^-}{2\pi}dy_1^-dy_2^-
e^{ixp^+y^- - ix_Lp^+(y_2^- - y_1^-)} \theta(-y_2^-)\theta(y^--y_1^-)
\nonumber \\
&\times&\langle A|\bar{\psi}_q(0)\frac{\gamma^+}{2}\psi_q(y^-)
\bar{\psi}_{q_i}(y_1^-)\frac{\gamma^+}{2}\psi_{q_i}(y_2^-)|A\rangle
\,
,\label{eq:qqbartrix-H}
\end{eqnarray}
with the antiquark $\bar q_i$ carrying momentum fraction $x_L$.
This two-parton correlation function is
always associated with double-hard rescattering
processes. Similarly, we define three other quark-antiquark
correlation matrix elements
\begin{eqnarray}
T^{A(S)}_{q{\bar q_i}}(x,x_L)&\equiv&\int\frac{p^+dy^-}{2\pi}dy_1^-dy_2^-
e^{i(x+x_L)p^+y^-} \theta(y^--y_1^-) \nonumber \\
&\times&\theta(-y_2^-)\langle A|\bar{\psi}_q(0)\frac{\gamma^+}{2}\psi_q(y^-)
\bar{\psi}_{q_i}(y_1^-)\frac{\gamma^+}{2}\psi_{q_i}(y_2^-)|A\rangle
\,
,\label{eq:qqbartrix-S} \\
T^{A(I-L)}_{q{\bar q_i}}(x,x_L)&\equiv&\int\frac{p^+dy^-}{2\pi}dy_1^-dy_2^-
e^{i(x+x_L)p^+y^- - ix_Lp^+(y^- - y_1^-)} \theta(y^--y_1^-)
\nonumber \\
&\times&\theta(-y_2^-) \langle A|\bar{\psi}_q(0)\frac{\gamma^+}{2}\psi_q(y^-)
\bar{\psi}_{q_i}(y_1^-)\frac{\gamma^+}{2}\psi_{q_i}(y_2^-)|A\rangle
\, ,
\label{eq:qqbartrix-I-L} \\
T^{A(I-R)}_{q{\bar q_i}}(x,x_L)&\equiv&\int\frac{p^+dy^-}{2\pi}dy_1^-dy_2^-
e^{i(x+x_L)p^+y^- - ix_Lp^+y_2^-}\theta(y^--y_1^-)
 \nonumber \\
&\times&\theta(-y_2^-)\langle A|\bar{\psi}_q(0)\frac{\gamma^+}{2}\psi_q(y^-)
\bar{\psi}_{q_i}(y_1^-)\frac{\gamma^+}{2}\psi_{q_i}(y_2^-)|A\rangle
\,
,\label{eq:qqbartrix-I-R}
\end{eqnarray}
that are associated with hard-soft rescattering and interference
between double hard and hard-soft rescattering.
In the first parton correlation
$T^{A(H)}_{q{\bar q_i}}(x,x_L)$, the antiquark $\bar q_i$
carries momentum fraction $x_L$ while the initial quark
has the momentum fraction $x$. The
two-parton correlation $T^{A(S)}_{q{\bar q_i}}(x,x_L)$
corresponds to the case when the leading quark has $x+x_L$
but the antiquark carries zero momentum.
The two interference matrix elements are approximately
the same for small value of $x_L$ and will be denoted as
$T^{A(I)}_{q{\bar q_i}}(x,x_L)$.

\subsection{$q\bar q \rightarrow q_i\bar q_i$ annihilation}

Contributions from the next-to-leading order quark-antiquark
annihilation or quark-quark (antiquark) scattering are more
complicated since they involve many real and virtual corrections.
The simplest real correction comes from $q\bar q\rightarrow q_i\bar
q_i$ annihilation ($q_i\neq q$) [Fig.~\ref{fig13} and
Eqs.~(\ref{eq:Ap-13}) and (\ref{eq:I13C})] which has
only a central-cut diagram,
\begin{eqnarray}
\Delta D_{q\rightarrow h}^{(q\bar q\rightarrow q_i\bar q_i)}(z_h)
&=&\frac{C_F}{N_C} \frac{\alpha_s^2 x_B}{Q^2} \int\frac{d\ell_T^2}{\ell_T^2}
\int_{z_h}^1\frac{dz}{z} [z^2+(1-z)^2] \nonumber \\
&\times& \sum_{q_i\neq q}\left[D_{q_i\to h}(z_h/z)
+D_{\bar q_i\to h}(z_h/z)\right]
\frac{T^{A(H)}_{q \bar q}(x,x_L)}{f^A_q(x)} \, .
\end{eqnarray}

This kind of $q\bar q$ annihilation is truly a hard processes
and thus requires the second antiquark to carry finite initial momentum
fraction $x_L$. Furthermore, there are no other interfering
processes.

\subsection{$qq_i(\bar q_i) \rightarrow qq_i(\bar q_i)$ scattering}

Contributions from non-identical quark-quark scattering
$q\bar q_i\rightarrow q\bar q_i$ ($q_i\neq q$) are a little complicated
because they involve all three cut diagrams (central, left and right)
[Eqs.~(\ref{eq:Ap-2C})-(\ref{eq:I2R})]. One can factor out
the $\theta$-functions in the hard parts according
to Eq.~(\ref{eq:contact}) and re-organize the phase factors
in each cut diagram,
\begin{eqnarray}
I_{\ref{fig2},C}&=&e^{i(x+x_L)p^+y^-}
(1-e^{-ix_Lp^+y_2^-})(1-e^{-ix_Lp^+(y^- - y_1^-)}) \nonumber \\
&=&e^{i(x+x_L)p^+y^-}[1-e^{-ix_Lp^+y_2^-}-e^{-ix_Lp^+(y^- - y_1^-)}
+e^{-ix_Lp^+(y^-+ y_2^- - y_1^-)}] \, ; \nonumber \\
I_{\ref{fig2},L}&=&e^{i(x+x_L)p^+y^-}
(1-e^{-ix_Lp^+(y^- - y_1^-)}) \nonumber \\
&=&e^{i(x+x_L)p^+y^-}[1-e^{-ix_Lp^+y_2^-}-e^{-ix_Lp^+(y^- - y_1^-)}
+e^{-ix_Lp^+y_2^-}] \, ;\nonumber \\
I_{\ref{fig2},R}&=&e^{i(x+x_L)p^+y^-}
(1-e^{-ix_Lp^+y_2^-}) \nonumber \\
&=&e^{i(x+x_L)p^+y^-}[1-e^{-ix_Lp^+y_2^-}-e^{-ix_Lp^+(y^- - y_1^-)}
+e^{-ix_Lp^+(y^- - y_1^-)}] \, ,
\end{eqnarray}
such that the first three terms in each amplitude are the same. These three
common phase factors will give rise to a contact contribution for
all similar hard parts from the three cut diagrams, which we
will neglect since they are not nuclear enhanced. The remaining
part will have the following phase factors,
\begin{eqnarray}
\overline{I}_{\ref{fig2}}&=&e^{i(x+x_L)p^+y^-}[
\theta(-y_2^-)\theta(y^- - y_1^-)e^{-ix_Lp^+(y^- +y_2^- - y_1^-)}
\nonumber \\
&&\hspace{-0.6in}-\theta(y_1^- - y_2^-)\theta(y^- - y_1^-)e^{-ix_Lp^+y_2^-}
-\theta(-y_2^-)\theta(y_2^- - y_1^-)e^{-ix_Lp^+(y^- - y_1^-)}] \, .
\end{eqnarray}
Note that the phase factors of the last two terms in the above
equation give identical contributions to the matrix elements
when intergated over $y_1^-$ and $y_2^-$ as they differ
only by the substitution $y_2^-\leftrightarrow y_1^--y^-$.
One therefore can combine them with
$\theta(-y_2^-)\theta(y^- - y_1^-)e^{-ix_Lp^+(y^- - y_1^-)}$
to form another contact contribution (path-ordered) which can be
neglected. The final effective phase factor is then
\begin{equation}
\overline{I}_{\ref{fig2}}=e^{ixp^+y^- -ix_Lp^+(y_2^- - y_1^-)}
(1-e^{ix_Lp^+y_2^-})\, .
\end{equation}
Using the above effective phase factor, one can obtain the effective
modification to the quark fragmentation function due to
quark-antiquark scattering, $q\bar q_i\rightarrow q\bar q_i$,
\begin{eqnarray}
\Delta D_{q\rightarrow h}^{(q\bar q_i\rightarrow q\bar q_i)}(z_h)
&=&\frac{C_F}{N_c}\frac{\alpha_s^2 x_B}{Q^2}\int\frac{d\ell_T^2}{\ell_T^2}
\int_{z_h}^1\frac{dz}{z} \sum_{\bar q_i\neq \bar q} \left\{
\left[D_{q\to h}(z_h/z) \frac{1+z^2}{(1-z)^2} \right. \right.\nonumber \\
&&\hspace{0.0in}+ \left. D_{g\to h}(z_h/z)
\frac{1+(1-z)^2}{z^2}\right]
\frac{T^{A(HI)}_{q\bar q_i}(x,x_L)}{f^A_q(x)} \nonumber \\
&& \hspace{-0.8in}
+\left. \left[D_{\bar q_i\to h}(z_h/z))-D_{g\to h}(z_h/z)\right]
\frac{1+(1-z)^2}{z^2} \frac{T^{A(HS)}_{q\bar q_i}(x,x_L)}{f^A_q(x)}
 \right\} \nonumber \\
&=&\frac{C_F}{N_c}\frac{\alpha_s^2 x_B}{Q^2}\int\frac{d\ell_T^2}{\ell_T^2}
\int_{z_h}^1\frac{dz}{z} \sum_{\bar q_i\neq \bar q} \left\{
\left[D_{q\to h}(z_h/z) \frac{1+z^2}{(1-z)^2}
\right. \right.\nonumber \\
&&\hspace{0.0in}+\left. D_{\bar q_i\to h}(z_h/z) \frac{1+(1-z)^2}{z^2}  \right]
\frac{T^{A(HI)}_{q\bar q_i}(x,x_L)}{f^A_q(x)} \nonumber \\
&& \hspace{-0.8in} + \left.
[D_{\bar q_i\to h}(z_h/z) - D_{g\to h}(z_h/z)]\frac{1+(1-z)^2}{z^2}
\frac{T^{A(SI)}_{q\bar q_i}(x,x_L)}{f^A_q(x)} \right\} \, ,
\label{eq:dd-qqbar1}
\end{eqnarray}
where three types of two-parton correlations are defined:
\begin{eqnarray}
T^{A(HI)}_{q\bar q_i}(x,x_L)&\equiv& T^{A(H)}_{q\bar q_i}(x,x_L)
-T^{A(I)}_{q\bar q_i}(x,x_L) \nonumber \\
&=&\int\frac{p^+dy^-}{2\pi}dy_1^-dy_2^-
e^{ixp^+y^- - ix_Lp^+(y_2^- - y_1^-)} (1-e^{ix_Lp^+y_2^-}) \nonumber \\
&&\hspace{-0.7in}\times\langle A|\bar{\psi}_q(0)\frac{\gamma^+}{2}\psi_q(y^-)
\bar{\psi}_{q_i}(y_1^-)\frac{\gamma^+}{2}\psi_{q_i}(y_2^-)|A\rangle
\theta(-y_2^-)\theta(y^--y_1^-) \,\,
,\label{eq:qqbarmatrix-1} \\
T^{A(SI)}_{q\bar q_i}(x,x_L)&\equiv& T^{A(S)}_{q\bar q_i}(x,x_L)
-T^{A(I)}_{q\bar q_i}(x,x_L) \nonumber \\
&=&\int\frac{p^+dy^-}{2\pi}dy_1^-dy_2^-
e^{i(x+x_L)p^+y^-} (1-e^{-ix_Lp^+y_2^-}) \nonumber \\
&&\hspace{-0.7in}\times\langle A|\bar{\psi}_q(0)\frac{\gamma^+}{2}\psi_q(y^-)
\bar{\psi}_{q_i}(y_1^-)\frac{\gamma^+}{2}\psi_{q_i}(y_2^-)|A\rangle
\theta(-y_2^-)\theta(y^--y_1^-) \,\,
,\label{eq:qqbarmatrix-2} \\
T^{A(HS)}_{q \bar q_i}(x,x_L)&\equiv& T^{A(HI)}_{q\bar q_i}(x,x_L)
+T^{A(SI)}_{q\bar q_i}(x,x_L)\nonumber \\
&=&\int\frac{p^+
dy^-}{2\pi}dy_1^-dy_2^- e^{i(x+x_L)p^+y^-} (1-e^{-ix_Lp^+y_2^-}) \nonumber \\
&\times&(1-e^{-ix_Lp^+(y^--y_1^-)})
\theta(-y_2^-)\theta(y^--y_1^-)\nonumber \\
&&\hspace{-0.7in}\times\langle A|\bar{\psi}_q(0)\frac{\gamma^+}{2}\psi_q(y^-)
\bar{\psi}_{q_i}(y_1^-)\frac{\gamma^+}{2}\psi_{q_i}(y_2^-)
|A\rangle
\, .
\label{eq:qqbarmatrix-3}
\end{eqnarray}

One can similarly obtain the modification of quark fragmentation
from non-identical quark-quark scattering  by
replacing $\bar q_i \rightarrow q_i$ in Eq.~(\ref{eq:dd-qqbar1}),
\begin{eqnarray}
\Delta D_{q\rightarrow h}^{(qq_i\rightarrow q q_i)}(z_h)
&=&\frac{C_F}{N_c}\frac{\alpha_s^2
x_B}{Q^2}\int\frac{d\ell_T^2}{\ell_T^2} \int_{z_h}^1\frac{dz}{z}
\sum_{ q_i\neq  q} \left\{ \left[ D_{q\to h}(z_h/z) \frac{1+z^2}{(1-z)^2}
\right.\right. \nonumber \\
&+& \left.
D_{q_i\to h}(z_h/z) \frac{1+(1-z)^2}{z^2} \right] \frac{T^{A(HI)}_{q
q_i}(x,x_L)}{f^A_q(x)} \nonumber \\
&&\hspace{-0.7in}+\left. [D_{q_i\to h}(z_h/z) - D_{g\to
h}(z_h/z)]\frac{1+(1-z)^2}{z^2}
\frac{T^{A(SI)}_{q q_i}(x,x_L)}{f^A_q(x)} \right\}  \, . \label{eq:dd-qq1}
\end{eqnarray}
The two-quark correlations, $T^{A(HI)}_{qq_i}(x,x_L)$ and
$T^{A(SI)}_{qq_i}(x,x_L)$ can be obtained  from $T^{A(HI)}_{q\bar q_i}(x,x_L)$
and $T^{A(SI)}_{q{\bar q_i}}(x,x_L)$, respectively, by making
the replacements
$\psi_{q_i}(y_2)\rightarrow \bar \psi_{q_i}(y_2)$ and
$\bar \psi_{q_i}(y_1) \rightarrow \psi_{q_i}(y_1)$ in
Eqs.~(\ref{eq:qqbarmatrix-1}) and (\ref{eq:qqbarmatrix-2}),
\begin{eqnarray}
T^{A(HI)}_{qq_i}(x,x_L)&\equiv&\int\frac{p^+dy^-}{2\pi}dy_1^-dy_2^-
e^{ixp^+y^- - ix_Lp^+(y_2^- - y_1^-)} (1-e^{ix_Lp^+y_2^-}) \nonumber \\
&&\hspace{-0.7in}\times\langle A|\bar{\psi}_q(0)\frac{\gamma^+}{2}\psi_q(y^-)
\bar{\psi}_{q_i}(y_2^-)\frac{\gamma^+}{2}\psi_{q_i}(y_1^-)|A\rangle
\theta(-y_2^-)\theta(y^--y_1^-) \,\,
,\label{eq:qqmatrix-1} \\
T^{A(SI)}_{qq_i}(x,x_L)&=&\int\frac{p^+
dy^-}{2\pi}dy_1^-dy_2^- e^{i(x+x_L)p^+y^-}
(1-e^{-ix_Lp^+y_2^-})  \nonumber \\
&&\hspace{-0.7in}\times\langle A|\bar{\psi}_q(0)\frac{\gamma^+}{2}\psi_q(y^-)
\bar{\psi}_{q_i}(y_2^-)\frac{\gamma^+}{2}\psi_{q_i}(y_1^-)
|A\rangle \theta(-y_2^-)\theta(y^--y_1^-) \, ,
\label{eq:qqmatrix-2}
\end{eqnarray}
and $T^{A(HS)}_{qq_i}(x,x_L)=T^{A(HI)}_{qq_i}(x,x_L)+T^{A(SI)}_{qq_i}(x,x_L)$.

Note that the contribution from fragmentation of quark $q_i$ or
antiquark $\bar q_i$ only comes from the central-cut diagram. This
contribution is positive and is proportional to
$T^{A(HI)}_{q{\bar q_i}}(x,x_L)+T^{A(SI)}_{q{\bar q_i}}(x,x_L)$,
containing all four terms: hard-soft,
double-hard and both interference terms . The
gluon fragmentation comes only from the single-triple interferences
(left and right-cut diagrams). Its contribution is therefore
negative and partially cancels the production of $q_i(\bar q_i)$
from the hard-soft rescattering. The cancellation is not complete
since the gluon and quark fragmentation functions are different. The
structure of this hard-soft rescattering (quark plus gluon) is very
similar to the lowest order result of $q\bar q\rightarrow g$ in
Eq.~(\ref{eq:dd-qq-g}). It contributes to the modification of the
effective fragmentation function but does not contribute to the
energy loss. The energy loss of the leading quark comes only from
double-hard rescattering, since the leading quark fragmentation
comes both from the central-cut and single-triple interferences,
and the single-triple interference terms cancel the effect of hard-soft
scattering for the leading fragmentation. Its net contribution is
therefore proportional to $T^{A(HI)}_{q{\bar q_i(\bar q_i)}}$.
Since the double-hard rescattering amounts to elastic $qq_i(\bar
q_i)$ scattering, the effective energy loss is essentially elastic
energy loss as shown in Ref.~\cite{Wang:2006qr}. There is, however,
LPM suppression due to partial cancellation by single-triple interference
contributions. For long formation time, $1/x_Lp^+ \gg R_A$, the
cancellation is complete. Therefore, LPM interference effectively
imposes the lower limit $x_L\ge 1/p^+R_A$ on the fractional momentum
carried by the second quark (antiquark).

\subsection{$qq\rightarrow qq$ scattering}

For identical quark-quark scattering, $qq\rightarrow qq$, one has to
include both $t$ and $u$-channels, their
interferences, and the related single-triple interference contributions.
Using the same technique to identify and neglect the contact contributions,
one can find the corresponding modification to the quark fragmentation
function from Eqs.~(\ref{eq:dd-qq1}) and (\ref{eq:Ap-21C})-(\ref{eq:I21R}),
\begin{eqnarray}
\Delta D_{q\rightarrow h}^{(qq\rightarrow q q)}(z_h)
&=&\frac{C_F}{N_c}\frac{\alpha_s^2 x_B}{Q^2}\int\frac{d\ell_T^2}{\ell_T^2}
\int_{z_h}^1\frac{dz}{z} \left\{\frac{T^{A(HS)}_{qq}(x,x_L)}{f^A_q(x)}
 \right. \nonumber \\
&&\hspace{-0.8in}\times \left[D_{q\to h}(z_h/z))-D_{g\to h}(z_h/z)\right]
\left(\frac{1+(1-z)^2}{z^2}-\frac{1}{N_c}\frac{1}{z(1-z)} \right) \nonumber \\
&&\hspace{-0.8in}+\left[D_{q\to h}(z_h/z) \left(\frac{1+z^2}{(1-z)^2}
-\frac{1}{N_c}\frac{1}{z(1-z)} \right) \right. \nonumber \\
&&\hspace{-0.8in}\left. \left.
+ D_{g\to h}(z_h/z) \left(\frac{1+(1-z)^2}{z^2}
-\frac{1}{N_c}\frac{1}{z(1-z)} \right)\right]
\frac{T^{A(HI)}_{qq}(x,x_L)}{f^A_q(x)} \right\} \nonumber \\
&&\hspace{-0.8in}=\frac{C_F}{N_c}\frac{\alpha_s^2 x_B}{Q^2}\int\frac{d\ell_T^2}{\ell_T^2}
\int_{z_h}^1\frac{dz}{z} \left\{\frac{T^{A(SI)}_{qq}(x,x_L)}{f^A_q(x)}
P_{qq\rightarrow qq}^{(s)}(z) [D_{q\to h}(z_h/z) \right. \nonumber \\
&&\hspace{-0.5in}-  D_{g\to h}(z_h/z)]
\left. + D_{q\to h}(z_h/z) P_{qq\rightarrow qq}(z)
\frac{T^{A(HI)}_{qq}(x,x_L)}{f^A_q(x)} \right\} \, ,
\label{eq:dd-qq2}
\end{eqnarray}
where the effective splitting functions are defined as
\begin{eqnarray}
P_{qq\rightarrow qq}^{(s)}(z)&=&
\frac{1+(1-z)^2}{z^2}-\frac{1}{N_c}\frac{1}{z(1-z)} \, ,
\label{eq:qqsplits}\\
P_{qq\rightarrow qq}(z)&=&
\frac{1+(1-z)^2}{z^2}+\frac{1+z^2}{(1-z)^2}-\frac{2}{N_c}\frac{1}{z(1-z)}
\, . \label{eq:qqsplit}
\end{eqnarray}

\subsection{$q\bar q \rightarrow q\bar q$, $gg$ annihilation}

The most complicated twist-four processes involving four quark
field operators
are quark-antiquark annihilation into two gluons or an identical
quark-antiquark pair. We have to consider them together since they
have similar single-triple interference processes and they involve the
same kind of quark-antiquark correlation matrix elements,
$T^{(i)}_{q\bar q}(x,x_L)$, $(i=HI,SI, HS)$.

For notation purpose, we first factor out the common factor
$(C_F/N_c)\alpha_s^2 x_B/Q^2/f^A_q(x)$ and the integration over
$\ell_T$ and $z$ and define
\begin{equation}
\Delta D_{q\rightarrow h}^{(q\bar q\rightarrow gg,q\bar q )}(z_h)
\equiv \frac{C_F}{N_c}\frac{\alpha_s^2 x_B}{Q^2f^A_q(x)}
\int\frac{d\ell_T^2}{\ell_T^2} \int_{z_h}^1\frac{dz}{z} \Delta
\widetilde{D}_{q\rightarrow h}^{(q\bar q\rightarrow gg,q\bar q )}
(z_h,z,x,x_L).
\end{equation}
After rearranging the
phase factors and identifying (by combining central, left and right
cut diagrams) and neglecting contact contributions
we can list in the following the twist-four corrections to the
quark fragmentation from the hard partonic parts of each cut
diagram (see Appendix A):

Fig.~\ref{fig1} ($t$-channel $q\bar q\rightarrow gg$),
\begin{eqnarray}
\Delta \widetilde{D}_{q\rightarrow h (\ref{fig1})}^{(q\bar q\rightarrow gg,q\bar q )}
&=&D_{g\to h}(z_h/z)2C_F\left[\frac{1+(1-z)^2}{z(1-z)}
+\frac{1+z^2}{z(1-z)}\right]T^{A(HI)}_{q\bar q}(x,x_L)
\nonumber \\
&&\hspace{-0.5in}+\left[D_{g\to h}(z_h/z)-D_{q\to h}(z_h/z)\right]
2C_F\frac{1+z^2}{z(1-z)}T^{A(SI)}_{q\bar q}(x,x_L)\, ;
\end{eqnarray}
Fig.~\ref{fig5}
(interference between $u$ and $t$-channel of $q\bar q\rightarrow gg$),
\begin{eqnarray}
\Delta \widetilde{D}_{q\rightarrow h (\ref{fig5})}^{(q\bar q\rightarrow gg,q\bar q )} &=&
D_{g\to h}(z_h/z)\frac{-4(C_F-C_A/2)}{z(1-z)}T^{A(HI)}_{q\bar q}(x,x_L)
\nonumber \\
&&\hspace{-0.5in}+\left[D_{g\to h}(z_h/z)-D_{q\to h}(z_h/z)\right]
\frac{-2(C_F-C_A/2)}{z(1-z)}T^{A(SI)}_{q\bar q}(x,x_L)\, ;
\end{eqnarray}
Fig.~\ref{fig14} ($s$-channel of $q\bar q \rightarrow gg$),
\begin{eqnarray}
\Delta \widetilde{D}_{q\rightarrow h (\ref{fig14})}^{(q\bar q\rightarrow gg,q\bar q )}=
D_{g\to h}(z_h/z)4C_A \frac{(1-z+z^2)^2}{z(1-z)}T^{A(H)}_{q\bar q}(x,x_L)
\, ;
\end{eqnarray}
Figs.~\ref{fig9} and \ref{fig10} (interference of $s$ and $t$-channel
$q \bar q\rightarrow g g$),
\begin{eqnarray}
\Delta \widetilde{D}_{q\rightarrow h (\ref{fig9}+\ref{fig10})}^{(q\bar q\rightarrow gg,q\bar q )}&=&
D_{g\to h}(z_h/z)(-2C_A)\left[\frac{1+z^3}{z(1-z)}
+ \frac{1+(1-z)^3}{z(1-z)} \right] \nonumber \\
&&\times T^{A(HI)}_{q\bar q}(x,x_L)
+ C_A\left[D_{q\to h}(z_h/z)\frac{1+z^3}{z(1-z)} \right. \nonumber \\
&&\hspace{-0.7in} \left. + D_{g\to h}(z_h/z)\frac{1+(1-z)^3}{z(1-z)}\right]
\times [T^{A(I2)}_{q\bar q}(x,x_L)-T^{A(I)}_{q\bar q}(x,x_L)]
\, ;
\end{eqnarray}
Fig.~\ref{fig13} ($s$-channel of $q\bar q\rightarrow q\bar q$),
\begin{eqnarray}
\Delta \widetilde{D}_{q\rightarrow h (\ref{fig13})}^{(q\bar q\rightarrow gg,q\bar q )}&=&
\left[D_{q\to h}(z_h/z)+D_{\bar q\to h}(z_h/z)\right]
[z^2+(1-z)^2]
 \nonumber \\
&& \hspace{2.0in}\times T^{A(H)}_{q\bar q}(x,x_L)\, ,
\end{eqnarray}
Fig.~\ref{fig2} ($t$-channel of $q\bar q\rightarrow q\bar q$), similar
to Eq.~(\ref{eq:dd-qqbar1}),
\begin{eqnarray}
\Delta \widetilde{D}_{q\rightarrow h (\ref{fig2})}^{(q\bar q\rightarrow gg,q\bar q )}&=&
D_{q\to h}(z_h/z) \frac{1+z^2}{(1-z)^2} T^{A(HI)}_{q\bar q}(x,x_L)
\nonumber \\
&& +D_{\bar q\to h}(z_h/z) \frac{1+(1-z)^2}{z^2} T^{A(HS)}_{q\bar q}(x,x_L)
\nonumber \\
&&- D_{g\to h}(z_h/z) \frac{1+(1-z)^2}{z^2}
T^{A(SI)}_{q\bar q}(x,x_L) \, ;
\end{eqnarray}
Figs.~\ref{fig3} and \ref{fig4} (interference between
$s$ and $t$-channel $q\bar q\rightarrow q\bar q$),
\begin{eqnarray}
\Delta \widetilde{D}_{q\rightarrow h (\ref{fig3}+\ref{fig4})}^{(q\bar q\rightarrow gg,q\bar q )}&=&
-4(C_F-C_A/2)\left[D_{q\to h}(z_h/z)\frac{z^2}{1-z} \right. \nonumber \\
&&\hspace{1.0in}+ \left. D_{\bar q\to h}(z_h/z)\frac{(1-z)^2}{z} \right]
T^{A(HI)}_{q\bar q}(x,x_L) \nonumber \\
&&\hspace{-0.6in}+2(C_F-C_A/2)\left[D_{q\to h}(z_h/z)\frac{z^2}{1-z}
+D_{g\to h}(z_h/z)\frac{(1-z)^2}{z} \right]  \nonumber \\
&\times&[T^{A(I2)}_{q\bar q}(x,x_L)-T^{A(I)}_{q\bar q}(x,x_L)]
\, ;
\end{eqnarray}
Figs.~\ref{fig7} and \ref{fig8} (two additional single-triple
interference diagrams),
\begin{eqnarray}
\Delta \widetilde{D}_{q\rightarrow h (\ref{fig7}+\ref{fig8})}^{(q\bar q\rightarrow gg,q\bar q )}&=&
-2C_F\left[D_{q\to h}(z_h/z)\frac{1+z^2}{1-z} \right. \nonumber \\
&&\hspace{0.5in}+ \left. D_{g\to h}(z_h/z)\frac{1+(1-z)^2}{z}\right]
T^{A(I2)}_{q\bar q}(x,x_L) \, .
\end{eqnarray}

Most processes involve
both $T^{A(HI)}(x,x_L)$ for double-hard rescattering with interference
and $T^{A(SI)}(x,x_L)$ for hard-soft rescattering with interference.
All the $s$-channel (Figs.~\ref{fig14} and \ref{fig13}) processes
involve double-hard scattering only. Therefore, they depend only on
the $T^{A(H)}_{q\bar q}(x,x_L)=T^{A(HI)}_{q\bar q}(x,x_L)
+T^{A(I)}_{q\bar q}(x,x_L)$. For
interference between single and triple scattering (left
and right-cut diagrams in Figs.~\ref{fig9}, \ref{fig10}, \ref{fig3}
\ref{fig4}, \ref{fig7} and \ref{fig8}), where a hard
rescattering with the second quark (antiquark) follows a soft
rescattering with the third antiquark (quark), only
interference matrix elements, $T^{A(I)}_{q\bar q}(x,x_L)$
and $T^{A(I2)}_{q\bar q}(x,x_L)$, are involved. Here,
\begin{eqnarray}
T^{A(I2)}_{q\bar q}(x,x_L)&\equiv&
\int\frac{p^+dy^-}{2\pi}dy_1^-dy_2^-
e^{ixp^+y^- + ix_Lp^+y_2^-} \nonumber \\
&\times&\langle A|\bar{\psi}_q(0)\frac{\gamma^+}{2}\psi_q(y^-)
\bar{\psi}_{q}(y_1^-)\frac{\gamma^+}{2}\psi_{q}(y_2^-)|A\rangle
\theta(-y_2^-)\theta(y^--y_1^-) \nonumber \\
&=&\int\frac{p^+dy^-}{2\pi}dy_1^-dy_2^-
e^{ixp^+y^- + ix_Lp^+(y^- - y_1^-)} \nonumber \\
&&\hspace{-0.5in} \times\langle A|\bar{\psi}_q(0)\frac{\gamma^+}{2}\psi_q(y^-)
\bar{\psi}_{q}(y_1^-)\frac{\gamma^+}{2}\psi_{q}(y_2^-)|A\rangle
\theta(-y_2^-)\theta(y^--y_1^-) \,\, ,\label{eq:qqbarI2}
\end{eqnarray}
is a new type of interference matrix elements that is only
associated with this type of single-triple interference
processes. One can categorize the above contributions according
to the associated two-quark correlation matrix elements
and rewrite the above contributions as,
\begin{eqnarray}
\Delta\widetilde{D}_{q\to h(HI)}^{q\bar q\to q\bar q, gg}
&=&T^{A(HI)}_{q\bar q}(x,x_L) [D_{g\to h}(z_h/z)P_{q\bar q\to gg}(z)
+D_{q\to h}(z_h/z)P_{q\bar q\to q\bar q}(z) \nonumber \\
&&\hspace{1.1in}
+D_{\bar q\to h}(z_h/z)P_{q\bar q\to q\bar q}(1-z)]
 \label{eq:qq-gg1}\\
\Delta\widetilde{D}_{q\to h(SI)}^{q\bar q\to q\bar q, gg}
&=&T^{A(SI)}_{q\bar q}(x,x_L)\left\{
\left[\frac{C_A}{z(1-z)}+2C_F\frac{z}{1-z}
-\frac{1+(1-z)^2}{z^2}\right]\right. \nonumber \\
&&\hspace{0in}\times D_{g\to h}(z_h/z)
-D_{q\to h}(z_h/z)\left[\frac{C_A}{z(1-z)}+2C_F\frac{z}{1-z}\right]
\nonumber \\
&& \hspace{1.0in}+\left. D_{\bar q\to h}(z_h/z)\frac{1+(1-z)^2}{z^2}\right\}
\nonumber \\
&=&T^{A(SI)}_{q\bar q}(x,x_L)\left\{\left[D_{q\to h}(z_h/z)
-D_{g\to h}(z_h/z)\right]
\left[P^{(s)}_{qq\to qq}(z) \right. \right. \nonumber \\
&& \hspace{-0.5in} -\left. 2C_F\frac{1+z^2}{z(1-z)}\right]
+\left. [D_{\bar q\to h}(z_h/z)-D_{q\to h}(z_h/z)]\frac{1+(1-z)^2}{z^2}
\right\}  \\
\Delta\widetilde{D}_{q\to h(I)}^{q\bar q\to q\bar q, gg}
&=&T^{A(I)}_{q\bar q}(x,x_L)
\left\{\left[C_A\frac{4(1-z+z^2)^2-1}{z(1-z)}
-2C_F\frac{(1-z)^2}{z}\right] \right. \nonumber \\
&\times&D_{g\to h}(z_h/z)+[z^2+(1-z)^2] D_{\bar q\to h}(z_h/z)]
\nonumber \\
&+&\left.D_{q\to h}(z_h/z)\left[z^2+(1-z)^2-\frac{C_A}{z(1-z)}-2C_F\frac{z^2}{1-z}
\right]+ \right\}
  \nonumber \\
&+&T^{A(I2)}_{q\bar q}(x,x_L)
\left\{D_{q\to h}(z_h/z)\left[\frac{C_A}{z(1-z)}-\frac{2C_F}{1-z}\right]
\right. \nonumber \\
&& \hspace{1.0in}
+\left. D_{g\to h}(z_h/z)\left[\frac{C_A}{z(1-z)}-\frac{2C_F}{z}\right]
\right\} \, ,
\end{eqnarray}
where $P^{(s)}_{qq\to qq}(z)$ is given in Eq.~(\ref{eq:qqsplits})
and the effective splitting functions for $q\bar q\rightarrow gg$
and $q\bar q\rightarrow q\bar q$ are defined as
\begin{eqnarray}
P_{q\bar q\rightarrow gg}(z)&=&
2C_F\frac{z^2+(1-z)^2}{z(1-z)}
-2C_A[z^2+(1-z)^2] \label{gg-split2} \,  ;\\
P_{q\bar q\rightarrow q\bar q}(z)&=&
z^2+(1-z)^2+\frac{1+z^2}{(1-z)^2}
  +\frac{2}{N_c}\frac{z^2}{1-z}\, ,
\label{qqbar-split2}
\end{eqnarray}
which come from the complete matrix elements of $q\bar q\rightarrow
gg$ and $q\bar q\rightarrow q\bar q$ scattering (see Appendix~\ref{appc}).
Again, double-hard rescattering corresponds to the elastic scattering
of the leading quark with another antiquark in the medium
and the interference contributions. The structure of the hard-soft
rescattering contribution we identify above shows the same kind of
gluon-quark (or quark-antiquark) mixing in the fragmentation
functions and does not
contribute to the energy loss of the leading quark. The unique
contributions in the $q\bar q\rightarrow q \bar q, gg$ processes are
the interference-only contributions. They mainly come
from single-triple interference processes in the multiple parton
scattering.

\section{Modification due to quark-gluon mixing}

We have so far cast the modification of the quark fragmentation
function due to quark-quark (antiquark) scattering (or annihilation)
in a form similar to the DGLAP evolution equation in vacuum. In fact,
one can also view the evolution of fragmentation functions in vacuum
as modification due to final-state gluon radiation. In both cases,
the modification
at large $z_h$ is mainly determined by the singular behavior of
the splitting functions for $z\rightarrow 1$, whereas the
modifications at mall $z_h$ is dominated by the singular behavior of the
splitting function for $z\rightarrow 0$.

Let us first focus on the modification at large $z_h$. A
careful examination of the contributions from all possible
processes shows that the dominant modification to the
effective quark fragmentation function comes from the $t$-channel of
double hard quark-quark scattering processes,
\begin{eqnarray}
\Delta D_{q\to h}(z_h)&\sim&\frac{C_F}{N_c}\frac{\alpha_s^2 x_B}{Q^2}
\sum_{q_i}\int \frac{d\ell_T^2}{\ell_T^2}
\int_{z_h}^1\frac{dz}{z} D_{q\to h}(\frac{z_h}{z})\left[\frac{T^{A(HI)}_{qq_i}(x,x_L)}{f^A_q(x)} \right. \nonumber \\
&&\hspace{1.5in} \times \frac{1+z^2}{(1-z)^2_+}\left. +\delta(1-z)\Delta_{q_i}(\ell_T^2)\right]
\nonumber \\
&=&\frac{C_F}{N_c}\alpha_s^2
\sum_{q_i}\int \frac{d\ell_T^2}{\ell_T^4}
\int_{z_h}^1\frac{dz}{z}  D_{q\to h}(\frac{z_h}{z})
\left[\frac{x_LT^{A(HI)}_{qq_i}(x,x_L)}{f^A_q(x)}
\right.\nonumber \\
&&\hspace{1.5in} \times \frac{z(1+z^2)}{(1-z)_+}+\left.\delta(1-z)\Delta_{q_i}(\ell_T^2)\right],
\end{eqnarray}
where the summation is over all possible quark and antiquark
flavors including $q_i=q, \bar q$ and $\Delta_{q_i}(\ell_T^2)$ represents
the contribution from virtual corrections. We have expressed
the modification in a form that it is proportional to the matrix
elements $x_LT^{A(HI)}_{qq_i}(x,x_L)/f^A_q(x)\sim A^{1/3}x_Lf^N_{q_i}(x_L)$
as compared to the modification from quark-gluon scattering
where the corresponding matrix element [Eq.~(\ref{eq:qgmatrix})]
is $T^{A(HI)}_{qg}(x,x_L)/f^A_q(x)\sim A^{1/3}x_LG^N(x_L)$.
Here, $f_{q_i}^N(x)$ and $G^N(x)$ are quark and gluon distributions,
respectively, in a nucleon. This leading contribution to the
modification from quark-quark scattering is very similar in form
to that from quark-gluon scattering [see Eq.~(\ref{eq:dd-qg})].
However, it is smaller due to the different color factors $C_F/C_A=4/9$
and the different quark and gluon distributions,
$f_{q_i}^N(x_L)$ and $G^N(x_L)$ in a nucleon. Because of LPM
intereference, small angle scattering with long formation
time $\tau_f=1/x_Lp^+$ is suppressed, leading to a
minimum value of  $x_L \ge x_A=1/m_NR_A=0.043$ for a $Kr$ target.
For this value of $x_L$, the ratio
\begin{equation}
\frac{\sum_{q_i}f_{q_i}^N(x_L,Q^2)}{G^N(x_L,Q^2)}
\ge 1.40/1.85 \sim 0.75,
\end{equation}
at $Q^2=2$ GeV$^2$ according the CTEG4HJ
parameterization \cite{cteq}.
Therefore, one has to include the effect of quark-quark
scattering for a complete calculation of the total quark energy loss
and medium modification of quark fragmentation functions.

In a weakly coupled and fully equilibrated quark-gluon
plasma, quark to gluon number density ratio is
$\rho_q/\rho_g=n_f(3/2)N_c/(N_c^2-1)=9n_f/16$. An
asymptotically energetic jet in an infinitely large medium
actually probes the small $x=\langle q_T^2\rangle/2ET$ regime,
where quark-antiquark pairs and gluons are predominantly generated
by thermal gluons through pQCD evolution. In this ideal scenario
one expects $N_q/N_g\sim 1/4C_A=1/12$ and therefore can neglect
quark-quark scattering. The
modification of quark fragmentation function will be dominated
by quark-gluon rescattering. However, for moderate jet
energy $E\approx 20$ GeV and a finite medium $L\sim 5$ fm,
parton distributions in a quark-gluon plamsa are close to
the thermal distribution. In particular, if quark and gluon
production is dominated by non-perturbative pair production
from strong color fields in the initial stage of heavy-ion
collisions \cite{lappi}, the quark to gluon ratio is comparable
to the equilibrium value. In this case, we should take into
account the medium modification of the quark fragmentation
functions by quark-quark scattering.

An important double hard process in quark-quark (antiquark)
scattering is $q\bar q\rightarrow gg$ [Eq.~(\ref{eq:qq-gg1})].
In this process, the annihilation converts the initial quark into
two final gluons that subsequently fragment into hadrons.
This will lead to suppression of the leading hadrons not only
because of energy loss (energy carried away by the other
gluon) but also due to the softer behavior of gluon fragmentation
functions at large $z_h$. Even though the leading behavior
of the effective splitting function [Eq.~(\ref{gg-split2})]
\begin{equation}
P_{q\bar q\to gg}(z)\approx 2C_F\frac{z^2+(1-z)^2}{z(1-z)}
\end{equation}
is not as dominating as that of $t$-channel quark-quark
scattering, it is enhanced by a color factor $2C_F=8/3$.
One expects this to make a significant contribution to the
medium modification at intermediate $z_h$.

In high-energy heavy-ion collisions, the ratios of
initial production rates for valence quarks, gluons and
antiquarks vary with the transverse momentum $p_T$.
Gluon production rate dominates at low $p_T$ while the
fraction of valence quark jets increases at large $p_T$.
Quarks are more likely to fragment into protons than
antiprotons, while gluons fragment into protons and antiprotons
with equal probabilities. Therefore, the ratio of
large $p_T$ antiproton and proton yields in $p+p$ collisions
is smaller than 1 and decreases with $p_T$ as  the fraction of
 valence quark jets
increases. Since gluons are expected to lose more energy than
quark jets, one would naively expect to see the antiproton
to proton ratio $\bar p/p$ becomes smaller due to jet quenching.
However, if the quark-gluon conversion due to $q\bar q\rightarrow gg$
becomes important, one would expect that the fractions of quark and gluon
jets are modified toward their equilibrium values. The final
$\bar p/p$ ratio could be larger than or comparable to
that in $p+p$ collisions.
Such a scenario of quark-gluon conversion was recently considered
in Ref.~\cite{ko} via a master rate equation.

The mixing between quark and gluon jets also happens at the lowest
order of quark-antiquark annihilation as shown in Fig.~\ref{fig0}.
At NLO, all hard-soft quark-quark (antiquark) scattering processes
have this kind of mixing between quark and gluon fragmentation functions.
Their contributions generally have the form,
\begin{eqnarray}
&&\frac{C_F}{N_c}\frac{\alpha_s^2 x_B}{Q^2}
\sum_{q_i}\int \frac{d\ell_T^2}{\ell_T^2}
\int_{z_h}^1\frac{dz}{z} \left[
D_{q_i\to h}(\frac{z_h}{z})-D_{g\to h}(\frac{z_h}{z})\right]
\nonumber \\
&&\hspace{2.5in}
\times P_{qq_i\to qq_i}(z)\frac{T^{A(SI)}_{qq_i}(x,x_L)}{f^A_q(x)},
\end{eqnarray}
where again the summation over the quark flavor
includes $q_i=q,\bar q$.
This mixing does not occur on the probability but rather on
on the amplitude level since it involves interferences
between single and triple scattering. Therefore, this
contribution depends on the difference between gluon and quark
fragmentation functions [Eq.~(\ref{eq:dd-qq-g})] and can be
positive or negative in different region of $z_h$. Nevertheless,
they contribute to the modification of the effective quark
fragmentation function and the flavor dependence of the final
hadron spectra.

\section{Flavor dependence of the medium modified fragmentation}

Summing all contributions to quark-quark (antiquark) double scattering
as listed in Section~\ref{MFF},
we can express the total twist-four correction up to
$\cal{O}$$(\alpha_s^2)$ to the quark fragmentation function as
\begin{eqnarray}
\Delta D_{q\to h}(z_h)&=&
\frac{C_F}{N_c}2\pi\frac{\alpha_s x_B}{Q^2}
\left\{2\left[D_{g\to h}(z_h)-D_{q\to h}(z_h) \right]
\frac{T^{A(H)}_{q \bar q}(x,0)}{f^A_q(x)} \right. \nonumber \\
&&\hspace{-0.3in}\left. +\frac{\alpha_s}{2\pi}\int\frac{d\ell_T^2}{\ell_T^2}
\int_{z_h}^1\frac{dz}{z}\sum_{a,b,i}
D_{b\to h}(z_h/z)P^{(i)}_{qa\to b}(z)
\frac{T^{A(i)}_{qa}(x,x_L)}{f^A_q(x)}\right\}\, ,
\label{dd-total}
\end{eqnarray}
where the summation is over all possible $q+a\rightarrow b+X$
processes and all different matrix elements
$T^{A(i)}_{qa}(x,x_L)$ ($i=HI,SI,I,I2$), which will be four
basic matrix elements we will use. The effective splitting
functions $P^{(i)}_{qa\to b}(z)$ are listed in Appendix~\ref{appb}.
One should also include virtual corrections which can be
constructed from the real corrections through unitarity
constraints \cite{GW}.

Similarly, we can also write down the twist-four corrections
to antiquark fragmentation in a nuclear medium,
\begin{eqnarray}
\Delta D_{\bar q\to h}(z_h)&=&
\frac{C_F}{N_c}2\pi\frac{\alpha_s x_B}{Q^2}
\left\{2\left[D_{g\to h}(z_h)-D_{\bar q\to h}(z_h) \right]
\frac{T^{A(H)}_{\bar q q}(x,0)}{f^A_{\bar q}(x)} \right. \nonumber \\
&&\hspace{-0.3in}\left. +\frac{\alpha_s}{2\pi}\int\frac{d\ell_T^2}{\ell_T^2}
\int_{z_h}^1\frac{dz}{z}\sum_{a,b,i}
D_{b\to h}(z_h/z)P^{(i)}_{\bar qa\to b}(z)
\frac{T^{A(i)}_{\bar qa}(x,x_L)}{f^A_{\bar q}(x)}\right\}\, ,
\label{ddqbar-total}
\end{eqnarray}
where the matrix elements $T^{A(i)}_{\bar qa}(x,x_L)$ and the
effective splitting functions $P^{(i)}_{\bar qa\to b}(z)$
can be obtained from the corresponding ones for quarks.
Given a model for the two-quark correlation functions, one will
be able to use the above expressions to numerically evaluate
twist-four corrections to the quark (antiquark) fragmentation
functions.
In this paper, we will instead give a qualitative
estimate of the flavor dependence of the correction in DIS
off a large nucleus.

For the purpose of a qualitative estimate, one can assume that all the
twist-four two-quark correlation functions can be factorized, as has
been done in Refs. \cite{GW,ZW,LQS,OW},
\begin{eqnarray}
&&\int \frac{p^+dy^{-}}{2\pi}\, dy_1^-dy_2^-
e^{ix_1p^+y^-+ix_2p^+(y_1^--y_2^-)} \theta(-y_2^-)\theta(y^--y_1^-)
\nonumber \\
& & \hspace{1.5in}\times
 \langle A|\bar{\psi}_q(0)\frac{\gamma^+}{2}\psi_q(y^-)
\bar{\psi}_{q_i}(y_1^-)\frac{\gamma^+}{2}\psi_{q_i}(y_2^-)|A\rangle
\nonumber \\
& & \hspace{1.5in} \approx \frac{C}{x_A} f_q^A(x_1)\,  f_{\bar
q_i}^N(x_2)\, ,
\label{eq:t4matrix} \\
&& \int \frac{p^+dy^{-}}{2\pi}\, dy_1^-dy_2^-
 e^{ix_1p^+y^-+ix_2p^+(y_1^--y_2^-) \pm ix_Lp^+y_2^-}
\theta(-y_2^-)\theta(y^--y_1^-)
\nonumber  \\
& & \hspace{1.5in}\times \langle
A|\bar{\psi}_q(0)\frac{\gamma^+}{2}\psi_{\bar q}(y^-)
\bar{\psi}_{q_i}(y_1^-)\frac{\gamma^+}{2}\psi_{q_i}(y_2^-)|A\rangle
\nonumber \\
&& \hspace{1.5in} \approx \frac{C}{x_A} f_q^A(x_1)\, f_{\bar
q_i}^N(x_2)e^{-x_L^2/x_A^2}\, , \label{eq:off-mx}
\end{eqnarray}
where $x_A=1/m_N R_A$, $m_N$ is the nucleon mass, $R_A$ the
nucleus size, $f_{\bar q_i}^N(x_2)$ is the antiquark distribution
in a nucleon and $C$ is assumed to be a constant, parameterizing
the strength of two-parton correlations inside a nucleus. The
integration over the position of the antiquark $(y_1^-+y_2^-)/2$
in the twist-four two-quark correlation matrix elements gives
rise to the nuclear enhancement factor $1/x_A=m_NR_A=0.21A^{1/3}$.

We should note that we set $k_T = 0$ for the collinear expansion.
As a consequence, the secondary quark
field in the twist-four parton matrix elements will carry zero
momentum in the soft-hard process. Finite intrinsic
transverse momentum leads to higher-twist corrections.
If a subset of the higher-twist terms in the collinear
expansion can be resummed to restore the phase factors such
as $ e^{ix_Tp^+y^-}$, where $x_T\equiv \langle k_T^2\rangle/2p^+q^-z(1-z)$,
the soft quark fields in the parton matrix elements will carry a
finite fractional momentum $x_T$.

Under such an assumption of factorization, one can
obtain all the two-quark correlation matrix elements:
\begin{eqnarray}
T^{A(HI)}_{q\bar q_i}(x,x_L)&\approx&\frac{C}{x_A} f_q^A(x)\, f_{\bar
q_i}^N(x_L+x_T) [1-e^{-x_L^2/x_A^2}],
\label{T-sim-1} \\
T^{A(SI)}_{q\bar q_i}(x,x_L)&\approx&
\frac{C}{x_A} f_q^A(x+x_L)\, f_{\bar
q_i}^N(x_T) [1-e^{-x_L^2/x_A^2}],  \label{T-sim-2}  \\
T^{A(I)}_{q\bar q_i}(x,x_L)&\approx&T^{A(I2)}_{q\bar q_i}(x,x_L)
\approx \frac{C}{x_A} f_q^A(x+x_L)\, f_{\bar q_i}^N(x_T)e^{-x_L^2/x_A^2}
\nonumber \\
&\approx &\frac{C}{x_A} f_q^A(x)\, f_{\bar q_i}^N(x_L+x_T)e^{-x_L^2/x_A^2}.
 \label{T-sim-3}
\end{eqnarray}
In the last approximation, we have assumed $x_L\sim x_T \ll x$.
Similarly, one can obtain $T^{A(i)}_{qq_i}(x,x_L)$,
$T^{A(i)}_{\bar qq_i}(x,x_L)$ and $T^{A(i)}_{\bar q\bar q_i}(x,x_L)$.
With these forms of two-quark correlation matrix elements, we
can estimate the flavor dependence of the nuclear
modification to the quark (antiquark) fragmentation functions.

The lowest order corrections [$\cal{O}$$(\alpha_s)$] are very
simple
\begin{eqnarray}
\Delta D_{q\to h}^{(LO)}(z_h)\propto
 C A^{1/3}[D_{g\to h}(z_h)-D_{q\to h}(z_h)]
f_{\bar q}^N(x_T)\, , \label{eq:lo1} \\
\Delta D_{\bar q\to \bar h}^{(LO)}(z_h)\propto
 C A^{1/3} [D_{g\to \bar h}(z_h)
-D_{\bar q\to \bar h}(z_h)]
f_{q}^N(x_T)\, .
\label{eq:lo2}
\end{eqnarray}
We consider the dominant contribution from the
fragmentation of a quark (antiquark)
which is one of the valence quarks (antiquarks) of the final
particle $h$ (antiparticle $\bar h$). The gluon fragmentation
functions into $h$ and $\bar h$ are the same. For large $z_h$,
the gluon fragmentation function is always softer than the
valence quark (antiquark) fragmentation \cite{gluon}.
Therefore, the lowest order twist-four corrections are always
negative for large $z_h$,
leading to a suppression of the valence quark (antiquark)
fragmentation function, $D_{q_v\to h}(z_h)$
[$D_{\bar q_v\to \bar h}(z_h)$]. Consider those quarks that
are also valence quarks of a nucleon:
\begin{eqnarray}
n \,\,& = &\,\, udd \nonumber \\
p \,\,& = &\,\, uud \,\,  ,\bar p \,\, = \,\, \bar u \bar u \bar
d\,\,\, , \label{eq:proton} \\
K^+ \,\,& = &\,\, u\bar s  \,\,  ,K^- \,\, = \,\, \bar u s \,\,\, .
\label{eq:kaon}\\
\pi^+, \pi^0, \pi^- \,\,
& = & \,\, u\bar d \,\,\,  , (u\bar u \, - d\bar d\, )/\sqrt 2
\,\, \, , d\bar u \,\,\,  . \label{eq:pion}
\end{eqnarray}
One can find the following flavor dependence of the lowest
order twist-four corrections to the quark (antiquark)
fragmentation functions,
\begin{equation}
\frac{\Delta D_{\bar q_v\to \bar h}^{(LO)}(z_h)} {\Delta D_{q_v\to
h}^{(LO)}(z_h)}=\frac{-|\Delta D_{\bar q_v\to \bar h}^{(LO)}(z_h)|}
{-|\Delta D_{q_v\to h}^{(LO)}(z_h)|}=\frac{f_{q_v}^N(x_T)}{f_{\bar
q_v}(x_T)}>1\, ,
\end{equation}
or
\begin{equation}
\frac{R^{(LO)}_{\bar q_v\to \bar h}(z_h)} {R^{(LO)}_{q_v\to
h}(z_h)}= \frac{1+\Delta D_{\bar q_v\to \bar h}^{(LO)}(z_h)/D_{\bar
q_\to \bar h}(z_h)} {1+\Delta D_{q_v\to h}^{(LO)}(z_h)/D_{ q_\to
h}(z_h)}<1\, ,
\end{equation}
where $R^{(LO)}_{q_v\to h}$ is the corresponding leading
order suppression of the fragmentation function at large $z_h$
for proton (anti-proton) and $K^+$ ($K^-$). Since pions contain
both valence quark and antiquark, the suppression factors should
be similar for all pions. For $x_T\ge 0.043$, $u(x)/\bar u(x)\ge 3$
and $d(x)/\bar d(x)\ge 2$ \cite{cteq}. Therefore, the modification of
antiquark fragmentation functions due to quark-antiquark
annihilation is significantly larger than that of a quark.

The flavor dependence of the NLO results are more complicated
since they involve scattering with both quarks and antiquarks
in the medium. One can observe first that effective splitting
functions (or quark-quark scattering cross section) are the same
for the $t$-channel $qq^\prime\rightarrow qq^\prime$ and
$q\bar q^\prime\rightarrow q \bar q^\prime$ ($q^\prime\neq q$)
scatterings,
\begin{equation}
P^{(i)}_{qq^\prime\to b}(z)=P^{(i)}_{\bar qq^\prime\to b}(z)
=P^{(i)}_{q\bar q^\prime\to b}(z)=P^{(i)}_{\bar q \bar q^\prime\to b}(z)\, .
\end{equation}

For identical quark-quark scattering or quark-antiquark annihilation,
one can separate the $q\bar q$ annihilation splitting functions
(or cross sections) into singlet and non-singlet contributions by
singling out the $t$-channel contributions,
\begin{eqnarray}
P^{(i)}_{q\bar q\to b}(z)&\equiv& P^{(i)}_{q q\to b}(z)
+\Delta P^{N(i)}_{q\bar q\to b}(z), \\
P^{(i)}_{\bar q q\to b}(z)&\equiv& P^{(i)}_{\bar q \bar q\to b}(z)
+\Delta P^{N(i)}_{\bar q q\to b}(z).
\end{eqnarray}
These singlet contributions to the modified fragmentation
functions are,
\begin{eqnarray}
\Delta D^{S(NLO)}_{q\to h}(z_h)&\propto&
\frac{\alpha_s}{2\pi}A^{1/3}\int\frac{d\ell_T^2}{\ell_T^2}
\sum_{b,q^\prime,i} D_{b\to h}\otimes P^{(i)}_{qq^\prime\to b}(z_h)
\nonumber \\
&&\hspace{1.5in}\times
[f^N_{q^\prime}(x_T)+f^N_{\bar q^\prime}(x_T)]C^{(i)}\, ,\\
\Delta D^{S(NLO)}_{\bar q\to \bar h}(z_h)&\propto&
\frac{\alpha_s}{2\pi}A^{1/3}\int\frac{d\ell_T^2}{\ell_T^2}
\sum_{b,q^\prime,i} D_{\bar b\to \bar h}\otimes
P^{(i)}_{\bar q \bar q^\prime\to \bar b}(z_h)
\nonumber \\
&&\hspace{1.5in}\times
[f^N_{q^\prime}(x_T)+f^N_{\bar q^\prime}(x_T)]C^{(i)}\, ,
\end{eqnarray}
where the summation over $q^\prime$ now includes $q^\prime$=$q$
and $C^{(i)}(x_L)$ are flavor-independent functions determined
from Eqs.~(\ref{T-sim-1})-(\ref{T-sim-3}),
\begin{eqnarray}
C^{(HI)}&=&C^{(SI)}=C(x_L)(1-e^{-x_L^2/x_A^2}),\nonumber \\
C^{(I)}&=&C^{(I2)}=C(x_L)e^{-x_L^2/x_A^2}\, ,
\end{eqnarray}
and $C(x_L)$ is a common coefficient that is a function
of $x_L$. Using $P^{(i)}_{\bar q \bar q\to \bar b}(z)=P^{(i)}_{q q\to b}(z)$
, one can conclude that the singlet
contributions to the modified quark and antiquark fragmentation
functions are the same, $\Delta D^{S(NLO)}_{q\to h}(z_h)=\Delta
D^{S(NLO)}_{\bar q\to \bar h}(z_h)$.

The non-singlet contributions, mainly from $s$-channel and $s$-$t$
interferences, are,
\begin{eqnarray}
\Delta D^{N(NLO)}_{q\to h}(z_h)&\propto&
\frac{\alpha_s}{2\pi}A^{1/3}\int\frac{d\ell_T^2}{\ell_T^2}
\sum_{b,i} D_{b\to h}\otimes \Delta P^{N(i)}_{q\bar q\to b}(z_h)
f^N_{\bar q}(x_T)C^{(i)}\, ,\\
\Delta D^{N(NLO)}_{\bar q\to \bar h}(z_h)&\propto&
\frac{\alpha_s}{2\pi}A^{1/3}\int\frac{d\ell_T^2}{\ell_T^2}
\sum_{\bar b,i} D_{\bar b\to \bar h}\otimes
\Delta P^{N(i)}_{\bar q q\to \bar b}(z_h)f^N_{q}(x_T)C^{(i)}\, ,
\end{eqnarray}
where again
$\Delta P^{N(i)}_{q\bar q\to b}(z)=\Delta P^{N(i)}_{\bar q q\to \bar b}(z)$
due to crossing symmetry. We have listed all non-vanishing
nonsinglet splitting functions $\Delta P^{N(i)}_{q\bar q\to b}(z)$ in
Appendix~\ref{appb}.

We again consider the limit $z_h\rightarrow 1$. In this region
the convolution in the modified fragmentation function is dominated
by the large $z\rightarrow 1$ behavior of the effective splitting
functions. From the listed $\Delta P^{N(i)}_{q\bar q\to b}(z)$
in Appendix~\ref{appb}, we can obtain the leading contributions,
\begin{eqnarray}
\sum_i C^{(i)}\Delta P^{N(i)}_{q\bar q\to q}(z)&\approx&
-4C_F\frac{C(x_L)}{1-z} ,\nonumber \\
\sum_i C^{(i)}\Delta P^{N(i)}_{q\bar q\to g}(z)&\approx&
2\left[2C_F+C_F(1-e^{-x_L^2/x_A^2})+C_Ae^{-x_L^2/x_A^2}\right]
\frac{C(x_L)}{1-z},
\end{eqnarray}
where we have also neglected terms proportional to $1/N_c$. All
$\Delta P^{N(i)}_{q\bar q\to \bar q}(z)$ are non-leading in
the limit $z\rightarrow 1$ and therefore can be neglected.
With these leading contributions, the non-singlet modification
to the quark and antiquark fragmentation functions can be
estimated as
\begin{eqnarray}
\Delta D^{N(NLO)}_{q\to h}(z_h)&\propto&
\frac{\alpha_s}{\pi}A^{1/3}\int\frac{d\ell_T^2}{\ell_T^2}
\int_{z_h}^1\frac{dz}{z}\left\{
D_{g\to h}\left(\frac{z_h}{z}\right)
\left[\frac{C(x_L)}{(1-z)_+} \right. \right. \nonumber \\
&\times& \left.  \left(C_F(1-e^{-x_L^2/x_A^2})
+ C_Ae^{-x_L^2/x_A^2}\right)  +\delta
(1-z)\Delta_1(\ell_T)\right]  \nonumber \\
&+& \left[D_{g\to h}\left(\frac{z_h}{z}\right)
-D_{q\to h}\left(\frac{z_h}{z}\right)\right] \nonumber \\
& \times & \left.\left[2C_F\frac{C(x_L)}{(1-z)_+} +
\delta (1-z)\Delta_2(\ell_T)\right] \right\} f^N_{\bar q}(x_T)
\, ,  \\
\Delta D^{N(NLO)}_{\bar q\to \bar h}(z_h)&\propto&
\frac{\alpha_s}{\pi}A^{1/3}\int\frac{d\ell_T^2}{\ell_T^2}
\int_{z_h}^1\frac{dz}{z}\left\{
D_{g\to \bar h}\left(\frac{z_h}{z}\right)
\left[\frac{C(x_L)}{(1-z)_+} \right. \right. \nonumber \\
&\times& \left. \left(C_F(1-e^{-x_L^2/x_A^2})
 + C_Ae^{-x_L^2/x_A^2}\right)  +\delta
(1-z)\Delta_1(\ell_T)\right] \nonumber \\
&+& \left[D_{g\to \bar
h}\left(\frac{z_h}{z}\right)
-D_{\bar q\to \bar h}\left(\frac{z_h}{z}\right)\right] \nonumber \\
&& \times  \left.\left[2C_F\frac{C(x_L)}{(1-z)_+} +
\delta (1-z)\Delta_2(\ell_T)\right] \right\} f^N_{q}(x_T)
\, ,
\end{eqnarray}
where $\Delta_1(\ell_T)$ and $\Delta_2(\ell_T)$ are from virtual corrections,
\begin{eqnarray}
\Delta_1(\ell_T)&=&\int_0^1 \frac{dz}{1-z}\left\{C_FC(x_L)|_{z=1}
-[C_F(1-e^{-x_L^2/x_A^2}) \right. \nonumber \\
&&\hspace{2.0in}+\left. C_Ae^{-x_L^2/x_A^2}]C(x_L)\right\}\, ,
\\
\Delta_2(\ell_T)&=&\int_0^1 \frac{dz}{1-z}2C_F\left[C(x_L)|_{z=1}-C(x_L)\right].
\end{eqnarray}
Because of momentum conservation, $C(x_L)=0$ when
$x_L\rightarrow \infty$ for $z=1$. Therefore, the above
virtual corrections are always negative. At large $z_h$,
these virtual corrections dominate over the real ones.

There are two kinds of non-singlet contributions in the
expressions given above. One that is proportional to
gluon fragmentation functions is
due to quark-antiquark annihilation into gluons which then fragment.
The fragmenting gluon not only carries less energy than the initial
quark but also has a softer fragmentation function, leading
to suppression of the final leading hadrons. The second type
of contributions is proportional to $D_{g\to h}(z_h)-D_{q\to h}(z_h)$
and therefore
mixes quark and gluon fragmentation functions, similarly as
the lowest order quark-antiquark annihilation processes
[see Eqs.~(\ref{eq:lo1}) and (\ref{eq:lo2})]. Since a gluon
fragmentation function is softer than a quark one, the real corrections
from this type of processes are positive for small $z_h$ and
negative for large $z_h$. The virtual corrections
have just the opposite behavior. Therefore, the second type of contributions
will reduce the total net modification. For intermediate
values of $z_h$
where $2D_{g\rightarrow h}(z_h)>D_{q\rightarrow h}(z_h)$, the net effect
is still the suppression of the effective fragmentation functions
for leading hadrons.

Since $f^N_{q}(x_T)>f^N_{\bar  q}(x_T)$, we can conclude that
the LO and NLO combined non-singlet suppression for antiquark
fragmentation into valence hadrons is larger than that for
quark fragmentation into valence hadrons.
This qualitatively explains the flavor dependence of nuclear
suppression of leading hadrons in DIS off heavy nuclear targets
as measured by the
HERMES experiment \cite{hermes:2000,hermes:2003}. The ratio of
differential semi-inclusive cross sections for nucleus and
deuteron targets were used to study the nuclear suppression of the
fragmentation functions. It was observed that suppression of leading
anti-proton is stronger than for leading proton and $K^-$ suppression is
stronger than $K^+$. In the valence quark fragmentation picture,
the leading proton ($K^+$) is produced mainly from $u$, $d$ ($u$) quark
fragmentation while anti-protons come primarily
from $\bar u$, $\bar d$ ($\bar u$)
fragmentation. Therefore, HERMES data are consistent
with stronger suppression of antiquark fragmentation.

Since gluon bremsstrahlung and the singlet $qq_i(\bar q_i)$
scattering also suppress quark and antiquark fragmentation, but
independently of quark flavor, one has to include all the processes
in order to have a complete and quantitative numerical evaluation of
the flavor dependence of the nuclear modification of the quark
fragmentation functions. Furthermore, the NLO contributions are
proportional to $\alpha_s\ln(Q^2)/2\pi$. They are as important as
the lowest order correction for large values of $Q^2$. In principle,
one should resum these higher order corrections via solving a set of
coupled DGLAP evolution equations, including medium modification for
gluon fragmentation functions. The contributions from quark-quark
(antiquark) scattering derived in this paper will be an important
part of the complete dscription. Detailed numerical study of the
effect of quark-quark (antiquark) scattering will be possible only
after the completion of this complete description in the future.

\section{Summary}

Utilizing the generalized factorization framework for twist-four
processes we have studied the nuclear modification of quark and
antiquark fragmentation functions (FF) due to quark-quark
(antiquark) double
scattering in dense nuclear matter up to order $\cal O$$(\alpha_s^2)$.
We calculated and analyzed the complete set of all possible cut diagrams.
The results can be categorized into contributions from double-hard,
hard-soft processes and their interferences. The double-hard rescatterings
correspond to elastic scattering of the leading quark with
another medium quark. It requires the second quark to carry
a finite fractional momentum $x_L$. Therefore, the energy loss
of the leading quark through such processes can be identified
as elastic energy loss at order $\cal O$$(\alpha_s^2)$.
The quark energy loss and modification of quark fragmentation
functions are dominated by the $t$-channel of quark-quark (antiquark)
scattering and are shown to be similar to that caused by quark-gluon
scattering. The contribution from quark-quark scattering is
smaller than that from quark-gluon scattering by a factor
of $C_F/C_A$ times the ratio of quark and gluon distribution functions
in the medium. We have shown that such contributions are not negligible
for realistic kinematics and finite medium size.
The soft-hard rescatterings mix
gluon and quark scattering, in the same way as the lowest
order $q\bar q\rightarrow g$ processes. Such processes modifies
the final hadron spectra or effective fragmentation functions but
do not contribute to energy loss of the leading quark.
For $q\bar q\rightarrow q\bar q, gg$ processes, there also exist
pure interference contributions mainly coming from
single-triple-scattering interference.

With a simple model of a factorized two-quark correlation functions,
we further investigated the flavor dependence of the medium
modified quark fragmentation functions in a large nucleus. We
identified the flavor dependent part of the modification
and find that the nuclear modification for an antiquark
fragmentation into a valence hadron is larger than that of
a quark. This offers an qualitative
explanation for the flavor dependence of the leading hadron
suppression in semi-inclusive DIS off  nuclear targets as
observed by the HERMES experiment \cite{hermes:2000,hermes:2003}.

\section*{Acknowledgements}

The authors thank Jian-Wei Qiu and Enke Wang for helpful discussion.
This work was supported by NSFC under project No. 10405011, by MOE
of China under project IRT0624, by Alexander von Humboldt
Foundation, by BMBF, by the Director, Office of Energy Research,
Office of High Energy and Nuclear Physics, Divisions of Nuclear
Physics, of the U.S. Department of Energy under Contract No.
DE-AC02-05CH11231, and by the US NSF under Grant No. PHY-0457265,
the Welch Foundation under Grant No. A-1358.

\renewcommand{\theequation}{A-\arabic{equation}}
  \setcounter{equation}{0}  
\renewcommand{\thesection}{A-\arabic{section}}
  \setcounter{section}{0}  

\section{Hard partonic parts for quark-quark double scattering}  
\label{appa}

\begin{figure}
\begin{center}
\includegraphics[width=80mm]{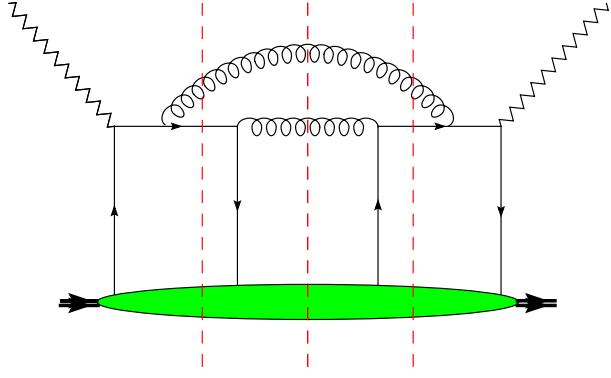}
\end{center}
\caption{The $t$-channel of $q\bar q\rightarrow gg$ annihilation
diagram with three possible cuts, central(C), left(L) and right(R).}
\label{fig1}
\end{figure}

\begin{figure}
\begin{center}
\includegraphics[width=80mm]{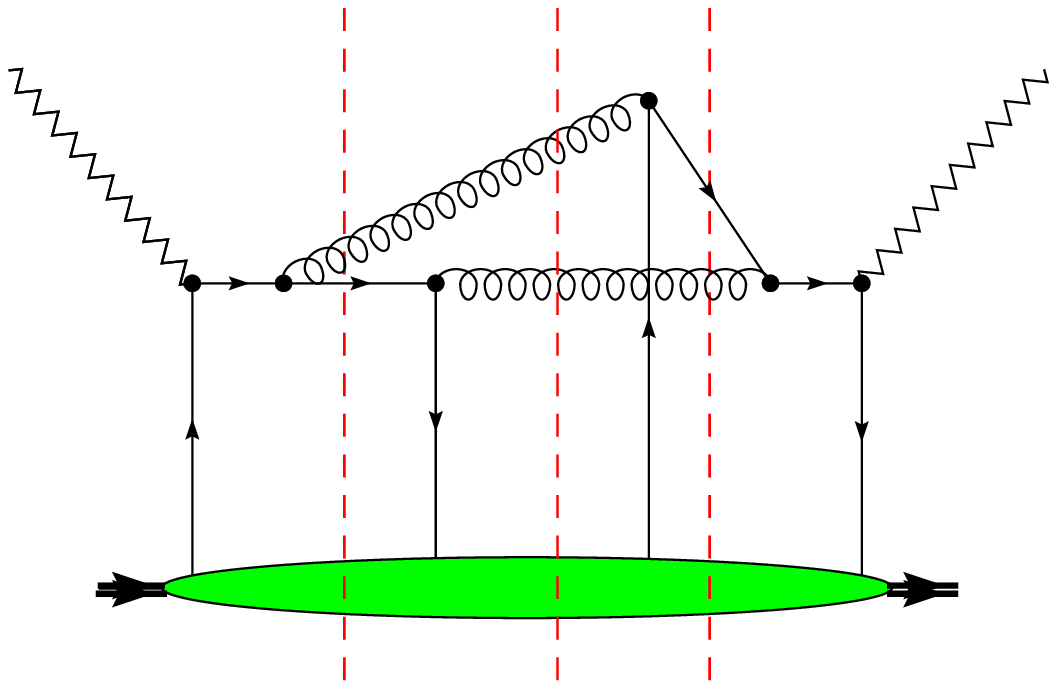}
\end{center}
\caption{The interference between $t$ and $u$-channel of
$q\bar q\rightarrow gg$ annihilation.} \label{fig5}
\end{figure}

\begin{figure}
\begin{center}
\includegraphics[width=80mm]{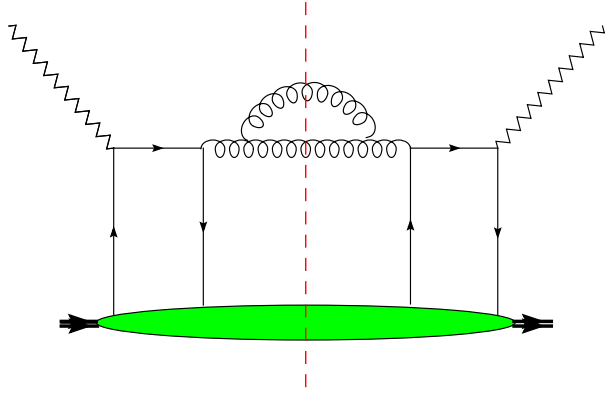}
\end{center}
\caption{The $s$-channel of $q\bar q\rightarrow gg$ annihilation
diagram with only a central-cut.} \label{fig14}
\end{figure}

\begin{figure}
\begin{center}
\includegraphics[width=80mm]{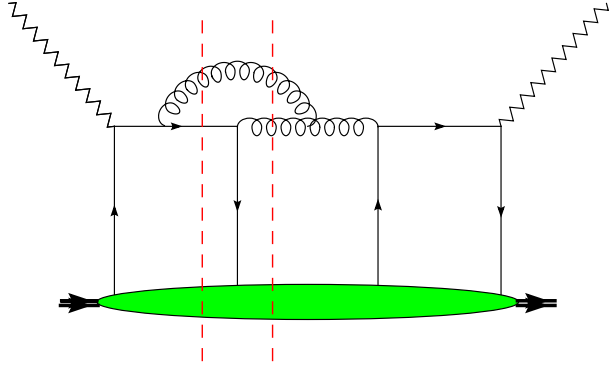}
\end{center}
\caption{The interference between $t$ and $s$-channel of
$q\bar q\rightarrow gg$ annihilation.} \label{fig9}
\end{figure}

\begin{figure}
\begin{center}
\includegraphics[width=80mm]{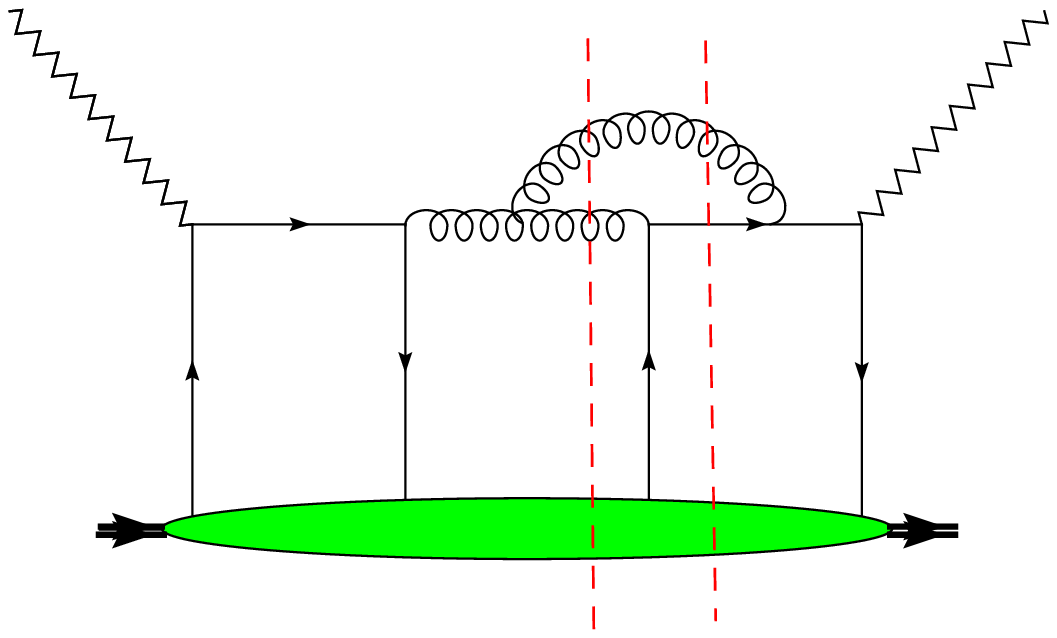}
\end{center}
\caption{The complex conjugate of  Fig.~\ref{fig9}. } \label{fig10}
\end{figure}

\begin{figure}
\begin{center}
\includegraphics[width=80mm]{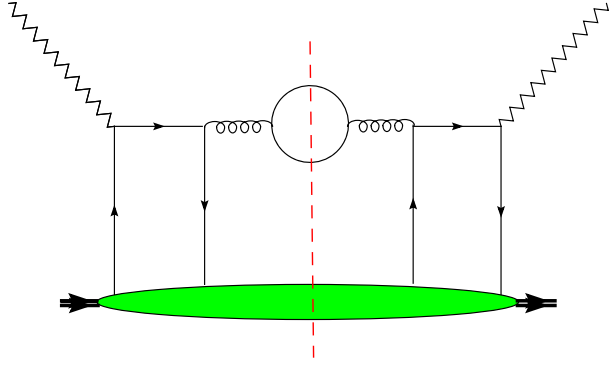}
\end{center}
\caption{$s$-channel $q\bar q \rightarrow q_i\bar q_i$
annihilation.} \label{fig13}
\end{figure}

\begin{figure}
\begin{center}
\includegraphics[width=80mm]{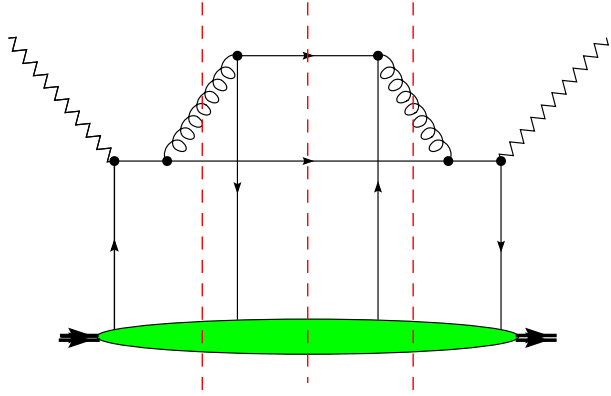}
\end{center}
\caption{$t$-channel $qq_i(\bar q_i) \rightarrow qq_i(\bar q_i)$
scattering.} \label{fig2}
\end{figure}

\begin{figure}
\begin{center}
\includegraphics[width=80mm]{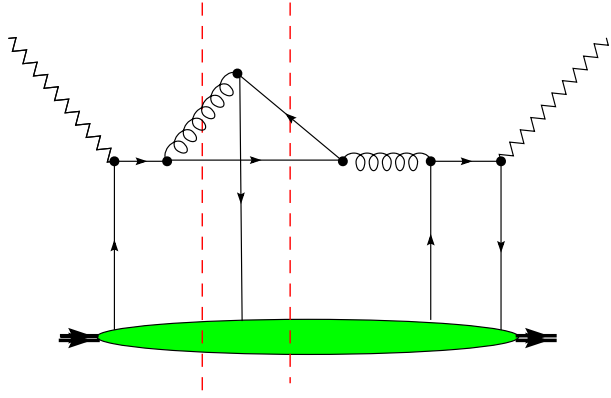}
\end{center}
\caption{Interference between $s$ and $t$-channel of
$q\bar q\rightarrow q\bar q$ scattering} \label{fig3}
\end{figure}

\begin{figure}
\begin{center}
\includegraphics[width=80mm]{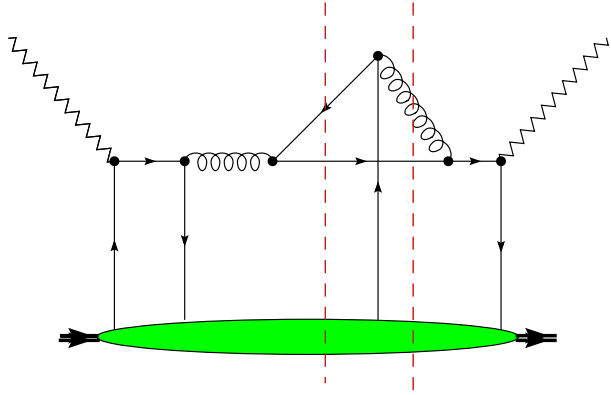}
\end{center}
\caption{The complex conjugate of Fig.~\ref{fig3}.} \label{fig4}
\end{figure}

\begin{figure}
\begin{center}
\includegraphics[width=80mm]{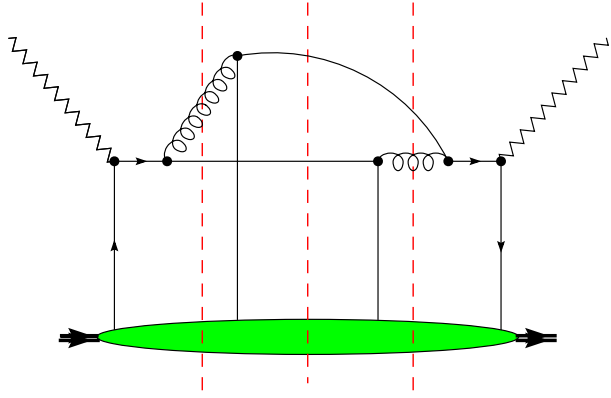}
\end{center}
\caption{The interference between $t$ and $u$-channel of
identical quark-quark scattering $qq \rightarrow qq$.} \label{fig21}
\end{figure}

\begin{figure}
\begin{center}
\includegraphics[width=80mm]{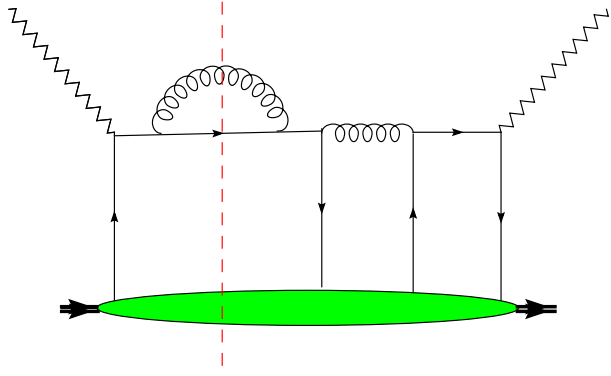}
\end{center}
\caption{Interference between final-state gluon radiation
from single and triple-quark scattering.} \label{fig7}
\end{figure}

\begin{figure}
\begin{center}
\includegraphics[width=80mm]{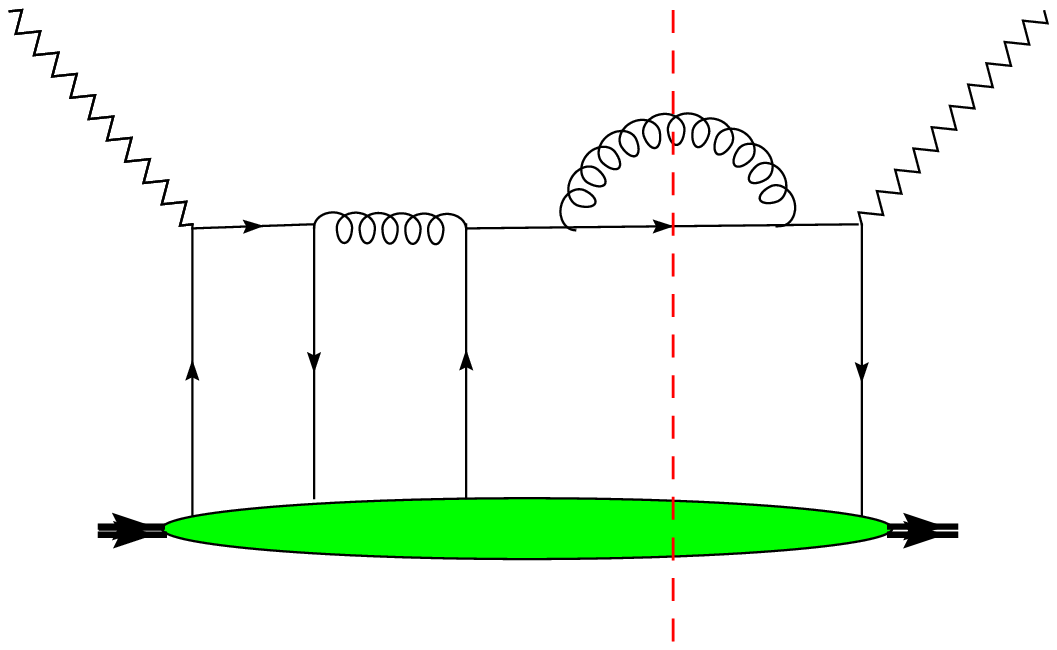}
\end{center}
\caption{The complex conjugate of Fig.~\ref{fig7}.} \label{fig8}
\end{figure}

In Section \ref{seciii} we have discussed the calculation of the hard part
of one example cut-diagram (Fig.~\ref{fig1}) in detail.
In this appendix we list the results for all possible
real corrections to quark-quark (antiquark) double scattering
in the next-to-leading order $\cal{O}$$(\alpha_s^2)$.
There are a total of 12 diagrams as
illustrated in Figs.~\ref{fig1}-\ref{fig8}.
For the purpose of abbreviation, we will suppress the
variables in the notations of partonic hard parts
\begin{equation}
\overline{H}^D\equiv
\overline{H}^D(y^-,y_1^-,y_2^-,x,p,q,z_h)\, ,
\end{equation}
and phase factor functions
\begin{equation}
\overline{I}\equiv
\overline{I}(y^-,y_1^-,y_2^-,x,,x_L,p)\, .
\end{equation}

We first consider all $q\bar q\rightarrow gg$ annihilation
diagrams with different possible cuts. The contributions of
Fig.~\ref{fig1} are:
\begin{eqnarray}
\overline{H}^D_{\ref{fig1},C}&=&
\frac{\alpha_s^2 x_B}{Q^2}\int\frac{d\ell_T^2}{\ell_T^2}
\int_{z_h}^1\frac{dz}{z}
\overline{I}_{\ref{fig1},C} D_{g\to h}(z_h/z) \nonumber \\
&&\hspace{1.2in}\times
\left[2\frac{1+z^2}{z(1-z)}+2\frac{1+(1-z)^2}{z(1-z)} \right]
\frac{C_F^2}{N_c}
 \, , \label{eq:Ap-1} \\
\overline{I}_{\ref{fig1},C}\,\,\,
&=&\theta(-y_2^-)\theta(y^- - y_1^-) e^{i(x+x_L)p^+y^-}\nonumber \\
&&\hspace{1.2in}\times(1-e^{-ix_Lp^+y_2^-})(1-e^{-ix_Lp^+(y^- - y_1^-)}) \, ,
\label{eq:I1C}
\end{eqnarray}
\begin{eqnarray}
\overline{H}^D_{\ref{fig1},L(R)}&=&
\frac{\alpha_s^2 x_B}{Q^2} \int\frac{d\ell_T^2}{\ell_T^2}
\int_{z_h}^1\frac{dz}{z}
\overline{I}_{\ref{fig1},L(R)} \,\,\,
\left[D_{q\to h}(z_h/z) 2\frac{1+z^2}{z(1-z)}\right.
\nonumber \\
&& \hspace{1.5in}\left. + D_{g\to h}(z_h/z) 2\frac{1+(1-z)^2}{z(1-z)}\right]
\frac{C_F^2}{N_c}
 \, , \label{eq:Ap-1-LR} \\
\overline{I}_{\ref{fig1},L}\,
&=&-\theta(y_1^- - y_2^-)\theta(y^- - y_1^-)
e^{i(x+x_L)p^+y^- }
(1-e^{-ix_Lp^+(y^- - y_1^-)}) \, ,\label{eq:I1L} \\
\overline{I}_{\ref{fig1},R}
&=&-\theta(-y_2^-)\theta(y_2^- - y_1^-)
e^{i(x+x_L)p^+y^-} (1-e^{-ix_Lp^+y_2^-}) \, .\label{eq:I1R}
\end{eqnarray}
Here we have included the fragmentation of both
final-state partons.


The contributions from Fig.~\ref{fig5} are:
\begin{eqnarray}
\overline{H}^D_{\ref{fig5},C} &=&
\frac{\alpha_s^2 x_B}{Q^2} \int\frac{d\ell_T^2}{\ell_T^2}
\int_{z_h}^1\frac{dz}{z}
\overline{I}_{\ref{fig5},C}\,\,\, 2 D_{g\to h}(z_h/z) \nonumber \\
&&\hspace{1.5in}\times \frac{-2}{(1-z)z}
 \frac{C_F(C_F-C_A/2)}{N_c}
 \, , \label{eq:Ap-5-1} \\
\overline{H}^D_{\ref{fig5},L(R)} &=&
\frac{\alpha_s^2 x_B}{Q^2} \int\frac{d\ell_T^2}{\ell_T^2}
\int_{z_h}^1\frac{dz}{z}
\overline{I}_{\ref{fig5},L(R)} \left[D_{g\to h}(z_h/z)
+D_{q\to h}(z_h/z)\right]\nonumber \\
&&\hspace{1.5in}\times \frac{-2}{(1-z)z}
 \frac{C_F(C_F-C_A/2)}{N_c}
 \, , \label{eq:Ap-5-2} \\
\overline{I}_{\ref{fig5},C}
&=&\theta(-y_2^-)\theta(y^- - y_1^-) e^{i(x+x_L)p^+y^-}
\nonumber \\
&\times &(1-e^{-ix_Lp^+y_2^-})(1-e^{-ix_Lp^+(y^- - y_1^-)}) \, ,
\label{eq:I5C}  \\
\overline{I}_{\ref{fig5},L} \,
&=&-\theta(y_1^- - y_2^-)\theta(y^- - y_1^-)
e^{i(x+x_L)p^+y^-} (1-e^{-ix_Lp^+(y^- - y_1^-)}) \, .\label{eq:I5L}\\
\overline{I}_{\ref{fig5},R}
&=&-\theta(-y_2^-)\theta(y_2^- - y_1^-)
e^{i(x+x_L)p^+y^-} (1-e^{-ix_Lp^+y_2^-}) \, ,\label{eq:I5R}
\end{eqnarray}
Note that the central-cut diagram in Fig.~\ref{fig5} corresponds
to the interference between $t$ and $u$-channel of the $q\bar q\rightarrow gg$
annihilation processes in Fig.~\ref{fig1}. Since the splitting function is
symmetric in $z$ and $1-z$, a factor of 2 comes from the fragmentation
of both gluons in the central-cut diagram.


The $s$-channel of $q\bar q \rightarrow gg$ is shown in Fig.~\ref{fig14}
which has only one central-cut. Its contribution to the partonic
hard part is,
\begin{eqnarray}
\overline{H}^D_{\ref{fig14},C} &=&
\frac{\alpha_s^2 x_B}{Q^2}\int\frac{d\ell_T^2}{\ell_T^2}
\int_{z_h}^1\frac{dz}{z}
\overline{I}_{\ref{fig14},C} \,\,
2D_{g\to h}(z_h/z)
 \frac{2(z^2-z+1)^2}{z(1-z)} \frac{C_FC_A}{N_c}
 \, , \label{eq:Ap-14} \\
\overline{I}_{\ref{fig14},C}
&=&\theta(-y_2^-)\theta(y^- - y_1^-) e^{i(x+x_L)p^+y^-}
e^{-ix_Lp^+(y^- - y_1^-)}e^{-ix_Lp^+y_2^-} \, .
\label{eq:I14C}
\end{eqnarray}
Note that the splitting function $2(z^2-z+1)^2/z(1-z)=2[1-z(1-z)]^2/z(1-z)$
is symmetric in $z$ and $1-z$. Therefore, fragmentation of the two
final gluons gives rise to the factor of 2 in front of the gluon
fragmentation function.


The interferences between $t$ and $s$-channel of $q\bar q \rightarrow gg$
processes are shown in Figs.~\ref{fig9} and \ref{fig10}. There
are only two possible cuts in these diagrams. The contributions
from Fig.~\ref{fig9} are:
\begin{eqnarray}
\overline{H}^D_{\ref{fig9},C} &=&
\frac{\alpha_s^2 x_B}{Q^2} \int\frac{d\ell_T^2}{\ell_T^2}
\int_{z_h}^1\frac{dz}{z}
\overline{I}_{\ref{fig9},C} \,\, D_{g\to h}(z_h/z)\nonumber \\
&&\hspace{1.2in}\times
 \left[2\frac{1+z^3}{z(1-z)} + 2\frac{1+(1-z)^3}{z(1-z)} \right]
\frac{C_FC_A}{2N_c}
 \, , \label{eq:Ap-9} \\
\overline{H}^D_{\ref{fig9},L} &=&
\frac{\alpha_s^2 x_B}{Q^2}\int\frac{d\ell_T^2}{\ell_T^2}
\int_{z_h}^1\frac{dz}{z}
\overline{I}_{\ref{fig9},L}
\left[D_{q\to h}(z_h/z)2\frac{1+z^3}{z(1-z)} \right. \nonumber \\
&& \hspace{1.2in}+ \left. D_{g\to h}(z_h/z)2\frac{1+(1-z)^3}{z(1-z)} \right]
\frac{C_FC_A}{2N_c}
 \, , \label{eq:Ap-9L} \\
\overline{I}_{\ref{fig9},C}
&=&\theta(-y_2^-)\theta(y^- - y_1^-)e^{i(x+x_L)p^+y^-}\nonumber \\
 &&\hspace{1.5in} \times
(1-e^{-ix_Lp^+y_2^-})e^{-ix_Lp^+(y^- - y_1^-)} \, , \\
\label{eq:I9C}
\overline{I}_{\ref{fig9},L} \,
&=&\theta(y_1^- - y_2^-)\theta(y^- - y_1^-)
e^{i(x+x_L)p^+y^- }\nonumber \\
&& \hspace{1.5in} \times (e^{-ix_Lp^+(y^- - y_2^-)}-e^{-ix_Lp^+(y^- - y_1^-)}) \, .
\label{eq:I9L}
\end{eqnarray}

Contributions from Fig.~\ref{fig10}, which are just the complex
conjugate of Fig.~\ref{fig9}, are:
\begin{eqnarray}
\overline{H}^D_{\ref{fig10},C} &=&
 \frac{\alpha_s^2 x_B}{Q^2}\int\frac{d\ell_T^2}{\ell_T^2}
\int_{z_h}^1\frac{dz}{z}
\overline{I}_{\ref{fig10},C} \,\, D_{g\to h}(z_h/z)\nonumber \\
&& \hspace{1.2in}\times
\left[2\frac{1+z^3}{z(1-z)}+2\frac{1+(1-z)^3}{z(1-z)}\right]
\frac{C_FC_A}{2N_c}
 \, , \label{eq:Ap-10} \\
\overline{H}^D_{\ref{fig10},R} &=&
\frac{\alpha_s^2 x_B}{Q^2} \int\frac{d\ell_T^2}{\ell_T^2}
\int_{z_h}^1\frac{dz}{z}
\overline{I}_{\ref{fig10},R} \left[D_{q\to h}(z_h/z)2\frac{1+z^3}{z(1-z)}
\right. \nonumber \\
&&\hspace{1.2in} + \left. D_{g\to h}(z_h/z)2\frac{1+(1-z)^3}{z(1-z)}\right]
\frac{C_FC_A}{2N_c}
 \, , \label{eq:Ap-10R} \\
\overline{I}_{\ref{fig10},C}
&=&\theta(-y_2^-)\theta(y^- - y_1^-)
e^{i(x+x_L)p^+y^-} \nonumber \\
& &\hspace{1.5in}\times(1-e^{-ix_Lp^+(y^- - y_1^-)})e^{-ix_Lp^+y_2^-} \, , \\
\label{eq:I10C}
\overline{I}_{\ref{fig10},R} \,
&=&\theta(-y_2^-)\theta(y_2^- - y_1^-) e^{i(x+x_L)p^+y^- }
\nonumber \\
&& \hspace{1.5in} \times(1-e^{-ix_Lp^+(y_2^- - y_1^-)})e^{-ix_Lp^+y_1^-} \, .
\label{eq:I10R}
\end{eqnarray}

One can collect all contributions of the double hard
$q\bar q \rightarrow g g$ processes from the central-cut
diagrams, which should have the common  phase factor
\begin{equation}
\bar{I}_C=\theta(-y_2^-)\theta(y^- - y_1^-)
e^{ixp^+y^-}e^{-ix_Lp^+(y_2^- - y_1^-)} \, ,
\end{equation}
and obtain the total effective splitting function in the
hard partonic part,
\begin{eqnarray}
\frac{C_F}{N_c}P_{q\bar q\rightarrow gg}(z)&=&\frac{2}{z(1-z)}\frac{1}{N_c}
\{ C_F^2 [1+z^2 + 1 +(1-z)^2]-2C_F(C_F-C_A/2) \nonumber \\
&+&2C_FC_A(1-z+z^2)^2-C_FC_A[1+z^3+1+(1-z)^3]\} \nonumber \\
&=&\frac{C_F}{N_c}\left[2C_F\frac{z^2+(1-z)^2}{z(1-z)}
-2C_A[z^2+(1-z)^2]\right]
\, . \label{gg-split}
\end{eqnarray}
We will find later in Appendix~\ref{appc} that the above result can also be
obtained from the total matrix elements squared for
$q\bar q \rightarrow g g$ annihilation.


We now consider the annihilation processes $q\bar q\rightarrow q_i \bar q_i$
with $q_i\neq q$. There is only the $s$-channel process with one
central-cut diagram as shown in Fig.~\ref{fig13}. Its contribution
to the hard part is
\begin{eqnarray}
\overline{H}^D_{\ref{fig13},C} &=&
\frac{\alpha_s^2 x_B}{Q^2} \int\frac{d\ell_T^2}{\ell_T^2}
\int_{z_h}^1\frac{dz}{z}
\overline{I}_{\ref{fig13},C}\sum_{q_i\neq q}\left[D_{q_i\to h}(z_h/z)
+D_{\bar q_i\to h}(z_h/z)\right] \nonumber \\
&&\hspace{2.0in}\times [z^2+(1-z)^2]
 \frac{C_F}{N_c}
 \, , \label{eq:Ap-13} \\
\overline{I}_{\ref{fig13},C}
&=& \theta(-y_2^-)\theta(y^- - y_1^-) e^{i(x+x_L)p^+y^-}
e^{-ix_Lp^+(y^- - y_1^-)}e^{-ix_Lp^+y_2^-} \, .
\label{eq:I13C}
\end{eqnarray}

Here we define the effective splitting function for
$q\bar q\rightarrow q_i \bar q_i$ annihilation as,
\begin{equation}
\frac{C_F}{N_c}P_{q\bar q\to q_i \bar q_i}(z)
=\frac{C_F}{N_c}[z^2+(1-z)^2] \, .
\end{equation}

Similarly, for $q \bar q_i \rightarrow q \bar q_i$
scattering with $q_i\neq q$, there is only the $t$-channel
as shown in Fig.~\ref{fig2}. There are, however, three cut
diagrams. Their contributions to the partonic hard part are:
\begin{eqnarray}
\overline{H}^D_{\ref{fig2},C} &=&
\frac{\alpha_s^2 x_B}{Q^2}\int\frac{d\ell_T^2}{\ell_T^2}
\int_{z_h}^1\frac{dz}{z}
\overline{I}_{\ref{fig2},C} \nonumber \\
&\times& \left[D_{q\to h}(z_h/z) \frac{1+z^2}{(1-z)^2}
+ D_{\bar q_i\to h}(z_h/z) \frac{1+(1-z)^2}{z^2}\right]
\frac{C_F}{N_c} \, , \label{eq:Ap-2C} \\
\overline{H}^D_{\ref{fig2},L(R)} &=&
\frac{\alpha_s^2 x_B}{Q^2}\int\frac{d\ell_T^2}{\ell_T^2}
\int_{z_h}^1\frac{dz}{z}
\overline{I}_{\ref{fig2},L(R)} \nonumber \\
&\times& \left[D_{q\to h}(z_h/z) \frac{1+z^2}{(1-z)^2}
+ D_{g\to h}(z_h/z) \frac{1+(1-z)^2}{z^2}\right]
\frac{C_F}{N_c} \, , \label{eq:Ap-2LR} \\
\overline{I}_{\ref{fig2},C}
&=&\theta(-y_2^-)\theta(y^- - y_1^-) e^{i(x+x_L)p^+y^-}\nonumber \\
&\times &(1-e^{-ix_Lp^+y_2^-})(1-e^{-ix_Lp^+(y^- - y_1^-)}) \, ,
\label{eq:I2C}  \\
\overline{I}_{\ref{fig2},L} \,
&=&-\theta(y_1^- - y_2^-)\theta(y^- - y_1^-)
e^{i(x+x_L)p^+y^-}
(1-e^{-ix_Lp^+(y^- - y_1^-)}) \, ,\label{eq:I2L} \\
\overline{I}_{\ref{fig2},R}
&=&-\theta(-y_2^-)\theta(y_2^- - y_1^-)e^{i(x+x_L)p^+y^-}
(1-e^{-ix_Lp^+y_2^-}) \, .\label{eq:I2R}
\end{eqnarray}
The  twist-four two-parton correlation
matrix element associated with the above quark-antiquark
scattering is the quark-antiquark correlator,
\begin{eqnarray}
T^A_{q\bar q_i}(x,x_L) &\propto&
e^{ixp^+y^- - ix_Lp^+(y_2^- - y_1^-)} \nonumber \\
&&\hspace{0.5in}\times\langle A|\bar{\psi}_q(0)\frac{\gamma^+}{2}\psi_q(y^-)
\bar{\psi}_{q_i}(y_1^-)\frac{\gamma^+}{2}\psi_{q_i}(y_2^-)|A\rangle\, ,
\end{eqnarray}
 and one should sum over all possible $q_i\neq q$ flavors. Note that
in the above matrix element, the momentum flow for the antiquark ($\bar q_i$)
is opposite to that of the quark ($q$) fields.

For quark-quark scattering, $qq_i\rightarrow qq_i$, the hard part
is essentially the same. The only difference is the associated
matrix element for the quark-quark correlator which is obtained
from that of the quark-antiquark correlator via the exchange
$\psi_{q_i}(y_2)\rightarrow \bar \psi_{q_i}(y_2)$ and
$\bar \psi_{q_i}(y_1) \rightarrow \psi_{q_i}(y_1)$,
\begin{eqnarray}
T^A_{q q_i}(x,x_L) &\propto&
e^{ixp^+y^- + ix_Lp^+(y_1^- - y_2^-)} \nonumber \\
&&\hspace{0.5in}\times
\langle A|\bar{\psi}_q(0)\frac{\gamma^+}{2}\psi_q(y^-)
\bar{\psi}_{q_i}(y_2^-)\frac{\gamma^+}{2}\psi_{q_i}(y_1^-)|A\rangle \, .
\end{eqnarray}
Note that the momentum flows of the two quarks ($q$ and $q_i$) point
in the same direction.

The effective splitting function of this scattering process is defined
through the fragmentation of the quark in the central-cut diagram,
\begin{equation}
\frac{C_F}{N_c}P_{qq_i(\bar q_i) \to qq_i(\bar q_i)}(z)=
\frac{C_F}{N_c}\frac{1+z^2}{(1-z)^2} \, .
\end{equation}

For annihilation $q\bar q\rightarrow q\bar q$ into identical quark
and antiquark pairs, in addition to the $s$-channel
(Fig.~\ref{fig13} for $q_i=q$) and $t$-channel (Fig.~\ref{fig2}
for $q_i=\bar q$), one has also to consider the interference
between $s$ and $t$-channel amplitudes as shown in Figs.~\ref{fig3}
and \ref{fig4}, each having two cuts. Their contributions to
the hard partonic parts are, respectively:
\begin{eqnarray}
\overline{H}^D_{\ref{fig3},C} &=&
\frac{\alpha_s^2 x_B}{Q^2} \int\frac{d\ell_T^2}{\ell_T^2}
\int_{z_h}^1\frac{dz}{z}
\overline{I}_{\ref{fig3},C} \left[D_{q\to h}(z_h/z)\frac{2z^2}{(1-z)}
\right. \nonumber \\
&&\hspace{0.8in}\left. + D_{\bar q\to h}(z_h/z)\frac{2(1-z)^2}{z}\right]
\frac{C_F(C_F-C_A/2)}{N_c}  \, , \label{eq:Ap-3C} \\
\overline{H}^D_{\ref{fig3},L} &=&
\frac{\alpha_s^2 x_B}{Q^2} \int\frac{d\ell_T^2}{\ell_T^2}
\int_{z_h}^1\frac{dz}{z}
\overline{I}_{\ref{fig3},L} \left[D_{q\to h}(z_h/z)\frac{2z^2}{(1-z)}
\right. \nonumber \\
&& \hspace{0.8in} \left. + D_{g\to h}(z_h/z)\frac{2(1-z)^2}{z}\right]
\frac{C_F(C_F-C_A/2)}{N_c}  \, , \label{eq:Ap-3L} \\
\overline{I}_{\ref{fig3},C}
&=& \theta(-y_2^-)\theta(y^- - y_1^-) e^{i(x+x_L)p^+y^-}
\nonumber \\
&& \hspace{0.8in}\times(1-e^{-ix_Lp^+y_2^-})e^{-ix_Lp^+(y^- - y_1^-)} \, ,
\label{eq:I3C}  \\
\overline{I}_{\ref{fig3},L}
&=&\theta(y_1^- - y_2^-)\theta(y^- - y_1^-) e^{i(x+x_L)p^+y^- }
\nonumber \\
&& \hspace{0.8in} \times(e^{-ix_Lp^+(y^- - y_2^-)}-e^{-ix_Lp^+(y^- - y_1^-)} ) \, ;
\label{eq:I3L}
\end{eqnarray}

\begin{eqnarray}
\overline{H}^D_{\ref{fig4},C} &=&
\frac{\alpha_s^2 x_B}{Q^2}\int\frac{d\ell_T^2}{\ell_T^2}
\int_{z_h}^1\frac{dz}{z}
\overline{I}_{\ref{fig4},C} \left[D_{q\to h}(z_h/z) \frac{2z^2}{(1-z)}
\right. \nonumber \\
 && \hspace{0.8in} \left. + D_{\bar q\to h}(z_h/z) \frac{2(1-z)^2}{z}\right]
\frac{C_F(C_F-C_A/2)}{N_c} \, , \label{eq:Ap-4C} \\
\overline{H}^D_{\ref{fig4},R} &=&
\frac{\alpha_s^2 x_B}{Q^2}\int\frac{d\ell_T^2}{\ell_T^2}
\int_{z_h}^1\frac{dz}{z}
\overline{I}_{\ref{fig4},R}  \left[D_{q\to h}(z_h/z) \frac{2z^2}{(1-z)}
\right. \nonumber \\
 &&\hspace{0.8in} \left. + D_{g\to h}(z_h/z) \frac{2(1-z)^2}{z}\right]
\frac{C_F(C_F-C_A/2)}{N_c} \, , \label{eq:Ap-4L} \\
\overline{I}_{\ref{fig4},C}
&=&\theta(-y_2^-)\theta(y^- - y_1^-) e^{i(x+x_L)p^+y^-} \nonumber \\
&& \hspace{0.8in} \times (1-e^{-ix_Lp^+(y^- - y_1^-)})e^{-ix_Lp^+y_2^-} \, ,
\label{eq:I4C}  \\
\overline{I}_{\ref{fig4},R} \,
&=& \theta(-y_2^-)\theta(y_2^- - y_1^-)e^{i(x+x_L)p^+y^- }
\nonumber \\
&& \hspace{0.8in} \times (e^{-ix_Lp^+y_1^-}-e^{-ix_Lp^+y_2^-}) \, . \label{eq:I4L}
\end{eqnarray}

One can again collect contributions from  the central-cut diagrams
of the double scattering processes in Figs.~\ref{fig13}, \ref{fig2}
\ref{fig3} and \ref{fig4} and obtain the total effective splitting
function for $q\bar q \rightarrow q\bar q$,
\begin{eqnarray}
\frac{C_F}{N_c}P_{q\bar q\rightarrow q\bar q}(z)&=&
\frac{C_F}{N_c}[z^2+(1-z)^2]+\frac{C_F}{N_c}\frac{1+z^2}{(1-z)^2}
-\frac{C_F(C_F-C_A/2)}{N_c}\frac{4z^2}{1-z} \nonumber \\
&=&\frac{C_F}{N_c}\left[ z^2+(1-z)^2+\frac{1+z^2}{(1-z)^2}
  +\frac{1}{N_c}\frac{2z^2}{1-z}\right] .
\label{qqbar-split}
\end{eqnarray}
Here we have used $C_F-C_A/2=-1/2N_c$. For antiquark fragmentation,
$P_{q\bar q\rightarrow \bar q q}(z)=P_{q\bar q\rightarrow q \bar q}(1-z)$.
One can also obtain the above result from $q\bar q \rightarrow q\bar q$
scattering matrix squared as shown in Appendix~\ref{appc}.


Similarly, for scattering of identical quarks $qq \rightarrow qq$,
one should set $q_i=q$ in Fig.~\ref{fig2}[in Eq.~(\ref{eq:Ap-2C})].
In addition, one should also also include interference
between $t$ and $u$-channel of the scattering as shown in Fig.~\ref{fig21}.
The contributions from such interference diagram are,
\begin{eqnarray}
\overline{H}^D_{\ref{fig21},C} &=&
\frac{\alpha_s^2 x_B}{Q^2}\int\frac{d\ell_T^2}{\ell_T^2}
\int_{z_h}^1\frac{dz}{z}
\overline{I}_{\ref{fig21},C}  \nonumber \\
&\times& 2 D_{q\to h}(z_h/z) \frac{2}{z(1-z)}
\frac{C_F(C_F-C_A/2)}{N_c} \, , \label{eq:Ap-21C} \\
\overline{H}^D_{\ref{fig21},L(R)} &=&
\frac{\alpha_s^2 x_B}{Q^2}\int\frac{d\ell_T^2}{\ell_T^2}
\int_{z_h}^1\frac{dz}{z}
\overline{I}_{\ref{fig21},L(R)}  \nonumber \\
&\times& \left[D_{q\to h}(z_h/z)+D_{g\to h}(z_h/z)\right]
 \frac{2}{z(1-z)} \frac{C_F(C_F-C_A/2)}{N_c} \, , \label{eq:Ap-21LR} \\
\overline{I}_{\ref{fig21},C}
&=&\theta(-y_2^-)\theta(y^- - y_1^-)
e^{i(x+x_L)p^+y^-} \nonumber \\
&\times& (1-e^{-ix_Lp^+y_2^-})(1-e^{-ix_Lp^+(y^- - y_1^-)}) \, ,
\label{eq:I21C}  \\
\overline{I}_{\ref{fig21},L} \,
&=&- \theta(y_1^- - y_2^-)\theta(y^- - y_1^-) e^{i(x+x_L)p^+y^-}
(1-e^{-ix_Lp^+(y^- - y_1^-)}) \, ,\label{eq:I21L} \\
\overline{I}_{\ref{fig21},R}
&=&- \theta(-y_2^-)\theta(y_2^- - y_1^-) e^{i(x+x_L)p^+y^-}
(1-e^{-ix_Lp^+y_2^-}) \, .\label{eq:I21R}
\end{eqnarray}
Note again that the fragmentation of both quarks contributes to
the factor 2 in Eq.~(\ref{eq:Ap-21C}) since the splitting function
is symmetric in $z$ and $1-z$. The twist-four two-quark correlation
matrix element associated with $qq\rightarrow qq$ scattering is
$T^A_{qq}(x,x_L)$ as compared to $T^A_{q\bar q}(x,x_L)$ for
quark-antiquark annihilation processes.

We can sum contributions from the double hard scattering
in all the central-cut diagrams in Figs.~\ref{fig2} and \ref{fig21}
and obtain the total effective splitting function for
$qq\rightarrow qq$ processes,
\begin{eqnarray}
\frac{C_F}{N_c}
P_{qq\rightarrow qq}(z)&=&\frac{C_F}{N_c}\left[\frac{1+z^2}{(1-z)^2}
+\frac{1+(1-z)^2}{z^2}\right]
+\frac{C_F(C_F-C_A/2)}{N_c}\frac{4}{z(1-z)} \nonumber \\
&=&\frac{C_F}{N_c}\left[\frac{1+z^2}{(1-z)^2}
+\frac{1+(1-z)^2}{z^2}-\frac{1}{N_c}\frac{2}{z(1-z)}\right] \, .
\label{qq-split}
\end{eqnarray}

There are two remaining cut diagrams that contribute to the
quark-antiquark  annihilation at the order of $\cal{O}$$(\alpha_s^2)$
as shown in Figs.~\ref{fig7} and \ref{fig8}. Their contributions
are:
\begin{eqnarray}
\overline{H}^D_{\ref{fig7},L} &=&
\frac{\alpha_s^2 x_B}{Q^2}\int\frac{d\ell_T^2}{\ell_T^2}
\int_{z_h}^1\frac{dz}{z}
\overline{I}_{\ref{fig7},L}  \nonumber \\
&\times&\left[ D_{q\to h}(z_h/z)2\frac{1+z^2}{1-z}
+ D_{g\to h}(z_h/z)2\frac{1+(1-z)^2}{z}\right]
\frac{C_F^2}{N_c} \, , \label{eq:Ap-7} \\
\overline{I}_{\ref{fig7},L}
&=&- \theta(y_1^- - y_2^-)\theta(y^- - y_1^-)e^{i(x+x_L)p^+y^-}
e^{-ix_Lp^+(y^- - y_2^-)} \, ,
\label{eq:I7}
\end{eqnarray}

\begin{eqnarray}
\overline{H}^D_{\ref{fig8},R} &=&
\frac{\alpha_s^2 x_B}{Q^2}\int\frac{d\ell_T^2}{\ell_T^2}
\int_{z_h}^1\frac{dz}{z}
\overline{I}_{\ref{fig8},R}  \nonumber \\
&\times& \left[D_{q\to h}(z_h/z)2\frac{1+z^2}{1-z}
+ D_{g\to h}(z_h/z)2\frac{1+(1-z)^2}{z}\right]
\frac{C_F^2}{N_c}  \, , \label{eq:Ap-8} \\
\overline{I}_{\ref{fig8},R} \,
&=&- \theta(-y_2^-)\theta(y_2^- - y_1^-) e^{i(x+x_L)p^+y^-}
e^{-ix_Lp^+y_1^-} \, . \label{eq:I8R}
\end{eqnarray}

\section{Effective splitting functions}
\label{appb}

In this Appendix, we list the effective splitting functions
associated with each process $qa\rightarrow b$ and the
double-hard ($HI$), hard-soft ($SI$) or their
interferences ($I,I2$) according to Eq.~(\ref{dd-total}).

\begin{eqnarray}
 P_{qq_i(\bar q_i)\to q_i(\bar q_i)}^{(HI)}(z)&=&\frac{1+(1-z)^2}{z^2}
 ,\,\,
 P_{qq_i(\bar q_i)\to q}^{(HI)}(z)=\frac{1+z^2}{(1-z)^2} ,\nonumber\\
P_{qq_i(\bar q_i)\to q_i(\bar q_i)}^{(SI)}(z)&=&\frac{1+(1-z)^2}{z^2}
 ,\,\, P_{qq_i(\bar q_i)\to g}^{(SI)}(z)=-\frac{1+(1-z)^2}{z^2} \label{spt1}\\
& & \nonumber\\
P_{q\bar q\to q_i}^{(HI)}(z)&=&P_{q\bar q\to \bar q_i}^{(HI)}(z)
=z^2+(1-z)^2 ,\,\, \nonumber \\
P_{q\bar q\to q_i}^{(I)}(z)&=&P_{q\bar q\to \bar q_i}^{(I)}(z)
=z^2+(1-z)^2 , \label{spt2} \\
& & \nonumber\\
P_{q q\to q}^{(HI)}(z)&=&\frac{1+(1-z)^2}{z^2}
+\frac{1+z^2}{(1-z)^2}-\frac{2}{N_c}\frac{1}{z(1-z)}\, , \,\,\, \nonumber\\
P_{q q\to g}^{(SI)}(z)&=&-P_{q q\to q}^{(SI)}(z)\, , \\
P_{q q\to q}^{(SI)}(z)&=&\frac{1+(1-z)^2}{z^2}
-\frac{1}{N_c}\frac{1}{z(1-z)} \, , \nonumber\\
& & \nonumber\\
P_{q \bar q\to q}^{(HI)}(z)&=&z^2+(1-z)^2+\frac{1+z^2}{(1-z)^2}
  +\frac{2}{N_c}\frac{z^2}{1-z}\, , \,\,\, \nonumber\\
P_{q \bar q\to \bar q}^{(HI)}(z)&=&P_{q \bar q\to q}^{(HI)}(1-z)
\, , \nonumber\\
P_{q \bar q\to g}^{(HI)}(z)&=&2C_F\frac{z^2+(1-z)^2}{z(1-z)}
-2C_A[z^2+(1-z)^2], \\
& & \nonumber\\
P_{q \bar q\to q}^{(SI)}(z)&=&-\left[\frac{C_A}{z(1-z)}
+2C_F\frac{z}{1-z}\right],\,\,\nonumber\\
P_{q \bar q\to \bar q}^{(SI)}(z)&=&\frac{1+(1-z)^2}{z^2},\nonumber\\
P_{q \bar q\to g}^{(SI)}(z)&=&\frac{C_A}{z(1-z)}+2C_F\frac{z}{1-z}
-\frac{1+(1-z)^2}{z^2} \\
& & \nonumber\\
P_{q \bar q\to q}^{(I)}(z)&=&z^2+(1-z)^2 -\frac{C_A}{z(1-z)}
-2C_F\frac{z^2}{1-z} ,\,\, \nonumber \\
P_{q \bar q\to \bar q}^{(I)}(z)&=&z^2+(1-z)^2 \, , \nonumber\\
P_{q \bar q\to g}^{(I)}(z)&=&C_A\frac{4(1-z+z^2)^2-1}{z(1-z)}
-2C_F\frac{(1-z)^2}{z} , \\
& & \nonumber\\
P_{q \bar q\to q}^{(I2)}(z)&=&\frac{C_A}{z(1-z)}-\frac{2C_F}{1-z}, \,\,
P_{q \bar q\to g}^{(I2)}(z)=\frac{C_A}{z(1-z)}-\frac{2C_F}{z}\, .
\end{eqnarray}

The non-singlet splitting functions for $q\bar q\rightarrow b$,
defined as
\begin{equation}
\Delta P_{q\bar q\to b}^{N(i)}(z)\equiv P_{q\bar q\to b}^{(i)}(z) -
P_{qq\to b}^{(i)}(z),
\end{equation}
are listed as below:
\begin{eqnarray}
\Delta P_{q\bar q\to q_i(\bar q_i)}^{N(HI)}(z)&=&
P_{q\bar q\to q_i(\bar q_i)}^{(HI)}(z),\,\,\,
\Delta P_{q\bar q\to q_i(\bar q_i)}^{N(I)}(z)=
P_{q\bar q\to q_i(\bar q_i)}^{(I)}(z), \\
& & \nonumber\\
\Delta P_{q\bar q\to q}^{N(HI)}(z)&=&-\frac{(1-z^2)(1+z^2+(1-z)^2)}{z^2}
+\frac{2}{N_c}\frac{1+z^3}{z(1-z)}, \nonumber \\
\Delta P_{q\bar q\to \bar q}^{N(HI)}(z)&=&P_{q \bar q\to \bar q}^{(HI)}(z)
,\,\,\,
\Delta P_{q\bar q\to g}^{N(HI)}(z)=P_{q \bar q\to g}^{(HI)}(z), \\
& & \nonumber\\
\Delta P_{q\bar q\to q}^{N(SI)}(z)&=&-\left[2C_F\frac{1+z^2}{1-z}
+\frac{1+(1-z)^2}{z^2}\right], \,\,\, \nonumber\\
\Delta P_{q\bar q\to \bar q}^{N(SI)}(z)&=&P_{q \bar q\to \bar q}^{(SI)}(z)
\,\nonumber \\
\Delta P_{q\bar q\to g}^{N(SI)}(z)&=&2C_F\frac{1+z^2}{1-z}
+\frac{2}{N_c}\frac{1}{z(1-z)}-2\frac{1+(1-z)^2}{z^2} \\
& & \nonumber\\
\Delta P_{q\bar q\to b}^{N(I)}(z)&=&P_{q\bar q\to b}^{(I)}(z),\,\,
\Delta P_{q\bar q\to b}^{N(I2)}(z)=P_{q\bar q\to b}^{(I2)}(z)\,
(b=q,\bar q, g)
\end{eqnarray}

\section{Alternative calculations of central-cut diagrams}
\label{appc}

As a cross-check of the hard partonic parts calculated from
different cut diagrams in Appendix~\ref{appa}, we provide an
alternative calculation of all the central-cut diagrams, which
correspond to quark-quark (antiquark) scattering.

Considering a parton ($a$) with momentum $q$ scattering with another
parton ($b)$ that carries a fractional momentum $xp$,
$a (q)+b(xp)\rightarrow c(\ell)+d(p^\prime)$, the cross section
can be written as
\begin{eqnarray}
d\sigma_{ab} &=&\frac{g^4}{2\hat s}
|M|^2_{ab\to cd}(\hat t/\hat s,\hat u/\hat s)
\frac{d^3\ell}{(2\pi)^3 2\ell_0}
2\pi \delta[(p+q-\ell)^2] \nonumber \\
&=&\frac{g^4}{(4\pi)^2}
|M|^2_{ab\to cd}(\hat t/\hat s,\hat u/\hat s)
\frac{\pi}{\hat{s}^2}
\frac{dz}{z(1-z)}d\ell_T^2 \,\, \delta\left(1-\frac{x_L}{x}\right),
\label{eq:cross-el}
\end{eqnarray}
where $q=[0,q^-,0]$ and $p=[xp^+,0,0]$ are momenta of the initial partons
and
\begin{equation}
\ell=\left[\frac{\ell_T^2}{2zq^-},zq^-, \vec{\ell}_T\right]
\end{equation}
is the momentum of one of the final partons. With the given kinematics,
the on-shell condition in the cross section can be recast as
\begin{eqnarray}
(xp+q-\ell)^2=2(1-z)xp^+q^-\left(1-\frac{x_L}{x}\right),\,\,\,\,\,
x_L=\frac{\ell_T^2}{2z(1-z)p^+q^-} .
\end{eqnarray}
The Mandelstam variables of the collision are,
\begin{eqnarray}
\hat s&=&(q+xp)^2=2xp^+q^-=\frac{\ell_T^2}{z(1-z)}, \,\,\,\,
\hat u=(\ell-xp)^2=-z\hat s,
\nonumber \\
\hat t&=&(\ell-q)^2=-(1-z)\frac{x_L}{x}\,\hat s=-(1-z)\hat s,
\label{eq:mand}
\end{eqnarray}
where we have used the on-shell condition $x=x_L$.

With Eq.~(\ref{eq:cross-el}) and parton distribution functions
$f^N_b(x)$, one can obtain the parton-nucleon cross section,
\begin{eqnarray}
d\sigma_{aN}&=&\sum _b d\sigma_{ab}f^N_b(x) dx \nonumber \\
&=&\sum _b  f^N_b(x_L)x_L
|M|^2_{ab\to cd}(\hat t/\hat s,\hat u/\hat s)
\frac{\pi\alpha_s^2}{\hat{s}^2}
\frac{dz}{z(1-z)}d\ell_T^2  \nonumber \\
&=& \sum _b  f^N_b(x_L) \frac{\pi\alpha_s^2}{s}C_0P_{ab\to cd}(z)dz
\frac{d\ell_T^2}{\ell_T^2}
\,\, ,
\label{eq:cross-el2}
\end{eqnarray}
where $s=2p^+q^-$ is the center-of-mass energy for $aN$ collision,
$C_0$ is some common color factor in the scattering matrix elements
and
\begin{equation}
P_{ab\to cd}(z)=(1/C_0)|M|^2_{ab\to cd}(\hat t/\hat s,\hat u/\hat s)
\end{equation}
is what we have defined as the effective splitting function for the
corresponding processes. One can therefore easily obtain these
effective splitting functions from the corresponding matrix
elements for elementary parton-parton scattering \cite{cross}.
We will list them in the following. A common color factor for all
quark-quark(antiquark) scattering is $C_0=C_F/N_c$.

$q\bar q \rightarrow q_i\bar q_i$ annihilation:
\begin{eqnarray}
|M|^2_{q\bar q\to q_i\bar q_i}&=&\frac{C_F}{N_c}
\frac{\hat{t}^2+\hat{u}^2}{\hat{s}^2}\,\, , \nonumber \\
P_{q\bar q\to q_i\bar q_i}(z)&=&z^2+(1-z)^2 \,\, .
\end{eqnarray}

$q\bar q \rightarrow q\bar q$ annihilation:
\begin{eqnarray}
|M|^2_{q\bar q \to q\bar q}&=&
\frac{C_F}{N_c}\left[\frac{\hat{u}^2+\hat{s}^2}{\hat{t}^2}
+\frac{\hat{u}^2+\hat{t}^2}{\hat{s}^2}
-\frac{1}{N_c}\frac{2\hat{u}^2}{\hat{s}\hat{t}} \right] , \nonumber \\
P_{q\bar q\to q \bar q}(z)&=&\frac{1+z^2}{(1-z)^2}+z^2+(1-z)^2
+\frac{2}{N_c}\frac{z^2}{1-z}\,\, .
\end{eqnarray}

$q\bar q \rightarrow gg$ annihilation:
\begin{eqnarray}
|M|^2_{q\bar q\to gg}&=&
\frac{C_F}{N_c}\left[2C_F\left(\frac{\hat{u}}{\hat{t}}+
\frac{\hat{t}}{\hat{u}}\right)
-2C_A\frac{\hat{u}^2+\hat{t}^2}{\hat{s}^2}\right], \nonumber \\
P_{q\bar q\to gg}(z)&=&2C_F\frac{z^2+(1-z)^2}{z(1-z)}
-2C_A(z^2+(1-z)^2)\, .
\end{eqnarray}

$qq_i(\bar q_i) \rightarrow q q_i(\bar q_i)$ scattering:
\begin{eqnarray}
|M|^2_{qq_i(\bar q_i)\to q q_i(\bar q_i)}&=&
\frac{C_F}{N_c}\frac{\hat{u}^2+\hat{s}^2}{\hat{t}^2} \nonumber \\
P_{qq_i(\bar q_i)\to q q_i(\bar q_i)}(z)&=&
\frac{1+z^2}{(1-z)^2}\,\,.
\end{eqnarray}

$qq \rightarrow q q$ scattering:
\begin{eqnarray}
|M|^2_{qq\to q q}&=&
\frac{C_F}{N_c}\left[\frac{\hat{u}^2+\hat{s}^2}{\hat{t}^2}
+\frac{\hat{s}^2+\hat{t}^2}{\hat{u}^2}
-\frac{1}{N_c}\frac{2\hat{s}^2}{\hat{t}\hat{u}} \right]\, , \nonumber \\
P_{qq \to q q}(z)
&=&\frac{1+z^2}{(1-z)^2}+\frac{1+(1-z)^2}{z^2}
-\frac{2}{N_c}\frac{1}{z(1-z)}\, .
\end{eqnarray}

For quark-gluon Compton scattering, the relevant gluon distribution
function is $x_LG_N(x_L)$. One can therefore rewrite contribution
from $qg\rightarrow qg$ to Eq.~(\ref{eq:cross-el2}) as,
\begin{eqnarray}
d\sigma_{qN}&=&x_LG_N(x_L)
\pi\alpha_s^2 z(1-z)|M|^2_{qg\to q g}(\hat t/\hat s,\hat u/\hat s)dz
\frac{d\ell_T^2}{\ell_T^4} \nonumber \\
&\equiv&x_LG_N(x_L)
\pi\alpha_s^2 \frac{C_F}{N_c}P_{qg\to qg}(z)dz
\frac{d\ell_T^2}{\ell_T^4}
\,\, .
\label{eq:cross-el3}
\end{eqnarray}

We have then for $qg\rightarrow qg$ scattering,
\begin{eqnarray}
|M|^2_{qg\to qg}&=&
\frac{C_A}{N_c}\frac{\hat{s}^2+\hat{u}^2}{\hat{t}^2}
-\frac{C_F}{N_c}\frac{\hat{u}^2+\hat{s}^2}{\hat{u}\hat{s}}
\nonumber \\
P_{qg \to qg}(z) &=&z(1-z)\left[\frac{C_A}{C_F}\frac{1+z^2}{(1-z)^2}
+\frac{1+z^2}{z}\right]\, .
\end{eqnarray}
Comparing this result with that in Ref.~\cite{GW} for the
quark-gluon rescattering, we can see that they agree in the limit
$1-z \rightarrow 0$. This is a consequence of the collinear
approximation employed in Ref.~\cite{GW} in the calculation of
the hard partonic part in quark-gluon rescattering.

We can also extend this calculation to the case of gluon-nucleon
scattering. One can use Eq.~(\ref{eq:cross-el2}) to define the splitting
function for $gq\rightarrow gq$ scattering,
\begin{eqnarray}
|M|^2_{gq\to gq}&=&
\frac{C_A}{N_c}\frac{\hat{s}^2+\hat{t}^2}{\hat{u}^2}
-\frac{C_F}{N_c}\frac{\hat{t}^2+\hat{s}^2}{\hat{t}\hat{s}}
\nonumber \\
P_{gq \to gq}(z) &=&z(1-z)\left[\frac{C_A}{N_c}\frac{1+(1-z)^2}{z^2}
+\frac{C_F}{N_c}\frac{1+(1-z)^2}{(1-z)}\right]\, .
\end{eqnarray}
Here for gluon-parton scattering, there is no common color factor.

$gg\rightarrow q\bar q$ annihilation,
\begin{eqnarray}
|M|^2_{gg\to q\bar q}&=&
\frac{1}{2N_c}\frac{\hat{t}^2+\hat{u}^2}{\hat{t}\hat{u}}
-\frac{1}{2C_F}\frac{\hat{t}^2+\hat{u}^2}{\hat{s}^2}
\nonumber \\
P_{gg \to q\bar q}(z)
&=&z(1-z)\left\{\frac{1}{2N_c}\frac{z^2+(1-z)^2}{z(1-z)}
-\frac{1}{2C_F}[z^2+(1-z)^2]\right\}\, .
\end{eqnarray}

$gg\rightarrow gg$ scattering
\begin{eqnarray}
|M|^2_{gg\to gg}&=&
2\frac{C_A}{C_F}\left[3-\frac{\hat{t}\hat{u}}{\hat{s}^2}
-\frac{\hat{u}\hat{s}}{\hat{t}^2} -\frac{\hat{t}\hat{s}}{\hat{u}^2}\right]
\nonumber \\
P_{gg \to gg}(z)
&=&2\frac{C_A}{C_F}\frac{(1-z+z^2)^3}{z(1-z)} \,\, .
\end{eqnarray}

One can use this technique to extend the study of modified fragmentation
functions to propagating gluons. Since the modification is dominated by
quark-gluon and gluon-gluon scattering, comparing the effective
splitting functions,
\begin{eqnarray}
\frac{C_F}{N_c}P_{qg\to qg}(z)&\approx& \frac{C_A}{N_c}\frac{2}{1-z}, \\
P_{gg\to gg}(z)&\approx& \frac{2C_A}{C_F}\frac{1}{1-z},
\end{eqnarray}
in the limit $z\rightarrow 1$, one can conclude that a gluon's
radiative energy loss is larger than a quark by a factor of
$N_c/C_F=C_A/C_F=9/4$. We will leave the complete derivation of medium
modification of gluon fragmentations to a future publication.



\end{document}